# Endo-exo classification of episodic rock creep in deep mines: Implications for forecasting catastrophic failure

Qinghua Lei[1], Daniel Francois Malan[2], Didier Sornette[3]

[1]Department of Earth Sciences, Uppsala University, Sweden

[2]Department of Mining Engineering, University of Pretoria, Hatfield, Pretoria, South Africa

[3]Institute of Risk Analysis, Prediction and Management, Academy for Advanced Interdisciplinary Studies, Southern University of Science and Technology, Shenzhen, China

Corresponding author: Qinghua Lei (qinghua.lei@geo.uu.se)

**Key Points:**

- An endo-exo theory explains episodic rock creep as a result of the interactions between exogenous triggers and endogenous deformations

- The theory classifies episodic rock creep into four distinct types, validated using closure measurements from a deep underground mine

- The theory is parsimonious with a single adjusted parameter accounting for all four power law regimes of episodic creep dynamics

**Abstract**

Rock masses in deep underground environments under high in-situ stress often exhibit episodic creep behavior, driven by complex interactions between external perturbation and internal reorganization. The causes of these creep episodes and their link to potential catastrophic failure remain poorly understood. Here, we present a novel "endo-exo" framework for analyzing episodic rock creep in deep underground mines, capturing the interplay between exogenous triggers (e.g., blasting and excavation) and endogenous processes (e.g., damage and healing within rock masses). The underlying physical mechanism involves cascades of locally triggered rock block movements due to fracturing and sliding. We identify four fundamental types of episodic dynamics, classified by the origin of disturbance (endogenous or exogenous) and the level of criticality (subcritical or critical). All four types exhibit power law relaxations with distinct exponents: $1+\theta$ (exogenous non-critical), $1-\theta$ (exogenous critical), $1-2\theta$ (endogenous critical) and $0$ (endogenous non-critical), all governed by a single parameter $0 < \theta < 1$. Our theoretical predictions are examined using the comprehensive dataset of a platinum mine in South Africa, where stopes display episodic closure behavior during successive mining operations. All creep episodes recorded can be accounted for in our classification with $\theta \approx 0.35\pm0.1$, providing strong validation of our theory. This $\theta$ value is interpreted in terms of a first-passage process driven by anomalous stress diffusion, represented by fractional Brownian motion or Lévy-type processes. Finally, we offer new insights into endo-exo interactions and the system's transition from episodic creep to catastrophic failure, with important implications for forecasting large-scale panel collapses.

**Plain Language Summary**

Amid ongoing efforts to reduce reliance on fossil fuels, there is a growing demand for critical minerals needed to build technologies like solar panels, wind turbines, and electric vehicles. To meet this demand, mining operations are going deeper underground, where the rocks are under high stresses. In these deep mines, rocks often creep in an episodic manner, sometimes leading to dangerous events known as rockbursts, which can damage infrastructure and endanger lives. However, the causes behind these episodic movements are still poorly understood. Are they triggered by external factors like blasting, by internal processes like damage accumulation, or both? Here, we introduce a novel endo-exo framework to improve our understanding of these events. This approach distinguishes between events originating from internal changes (endogenous) and those caused by outside forces (exogenous). Applying to data from a platinum mine in South Africa, we find that our theory successfully captures the full spectrum of observed creep events. Our framework reveals a deep quantitative link between episodic movements, external triggers, and internal processes in rock masses. The insights obtained not only help improve safety in deep mining but also offer a new way to understand similar processes in earthquakes, landslides, volcanoes, and glaciers.

## 1 Introduction

The ongoing green energy transition worldwide poses great new challenges for global supply chains of critical minerals that are essential for making solar panels, wind turbines and electric vehicle batteries. To achieve net-zero emissions by 2050, it is estimated that mineral inputs in 2040 must be six times greater than today's levels (International Energy Agency, 2021). However, as shallow mineral resources become increasingly depleted, mining must inevitably extend to greater depths. Deep mining operations face substantial challenges, especially rock mass instability under high in-situ stresses (Cook, 1965, 1976; Linkov, 1996; Ortlepp & Stacey, 1994; Ranjith et al., 2017). Extensive field observations reveal that rock masses surrounding deep mines commonly exhibit episodic creep behavior, characterized by sudden movements followed by gradual, time-dependent relaxation, typically triggered by mining or seismic activities (Hsiung et al., 1992; Malan, 1999; Malan et al., 2007; Malan & Napier, 2021; McGarr, 1971b, 1971a; McGarr et al., 1982; McGarr & Green, 1975). In some cases, rock masses undergo episodic creep for extended periods without collapsing, whereas in others, these episodic movements can escalate into violent rockbursts after hours to weeks of intermittent evolution (Lei & Sornette, 2025a, 2025b; Malan, 1999; Malan et al., 2007; McGarr, 1971a, 1971b; Ouillon & Sornette, 2000). The mechanisms behind these episodic creep responses and their link to potential catastrophic failure remain poorly understood, limiting our ability to forecast and mitigate violent rockburst events.

We identify three core research questions: (i) Does episodic rock creep have an exogenous or endogenous origin? (ii) What are the physical mechanisms driving these episodic movements? (iii) How do they contribute to or relate to violent rockbursts? In this context, we define exogenous (exo) triggers as external perturbations such as blasting operations, excavation activities, and seismic events occurring outside the domain of interest, while endogenous (endo) processes include damage accumulation, healing, internal faulting, and evolving frictional or material properties. To address these questions, we develop a novel "endo-exo" theoretical framework to quantitatively analyse the precursory and recovery patterns associated with episodic rock creep events. The rationale behind this is that complex systems like fractured rock masses composed of numerous interacting constituent components can display distinguishable precursory and recovery characteristics in response to large fluctuations (Sornette & Helmstetter, 2003). These signatures can be used to distinguish endogenous from exogenous origins of episodic dynamics— an approach that has been demonstrated for various geophysical phenomena such as earthquake sequences (Helmstetter et al., 2003; Helmstetter & Sornette, 2002) and landslide movements (Lei & Sornette, 2024). In fact, all real-world geophysical systems are fundamentally governed by the interplay of external perturbations and internal processes, rendering the endo-exo interaction a universal phenomenon (Sornette, 2006b). Examples include seismogenic faults (Scholz, 2019), volcanoes (Sahoo et al., 2024), landslides (Lacroix et al., 2020), glaciers (Giordan et al., 2020), and underground mines (Ortlepp & Stacey, 1994). These systems are out-of-equilibrium (i.e., with macroscopic properties evolving over time) and open (i.e., interacting with their surroundings). Our foundational endo-exo theory provides a comprehensive and quantitative framework for understanding the dynamical evolution of these diverse complex geophysical systems. It offers useful diagnostic tools for distinguishing endogenous factors (i.e., internal mechanisms governing system behavior) from exogenous influences (i.e., external shocks or interventions affecting the system), as well as entangled situations where strong interactions among system components produce intertwined causality. We emphasize that the broad applicability of our endo-exo theory stems not from a superficial analogy, but instead from the fundamental universality of the endo-exo concept and the generality of its mathematical formulation.

In this paper, with a specific focus on deep underground mines to test the endo-exo framework, we mainly study the episodic creep regime, where endogenous and exogenous origins can be distinguished, and only briefly investigate the catastrophic failure regime, which is covered more comprehensively in our previous work (Lei & Sornette, 2025b, 2025a; Ouillon & Sornette, 2000). We provide a detailed demonstration of the endo-exo framework using the stope closure monitoring data from a deep tabular platinum mine in South Africa. We further explore the physical mechanisms driving endo-exo interactions and the system's transition from episodic creep to catastrophic failure leading up to violent rockbursts. The insights obtained from the deep mine system, where underground excavations provide unique access to the source region of failure events, are not only relevant for mining safety, but are also of general and relevant interest in connection with earthquakes, landslides, volcanoes, and glaciers, where such direct access is limited or even impossible.

The remainder of the paper is organized as follows. Section 2 presents the endo-exo concept, mathematical formulations, and classification principles, as well as data acquisition and analysis methods. Section 3 reports the results, which are further discussed in Section 4. Finally, conclusions are drawn in Section 5.

**2 Methodology**

2.1 Problem conceptualization

Figure 1a shows the representative configuration of a tabular excavation in deep and intermediate-depth mines, where mining proceeds laterally by progressively advancing the stope faces through drilling and blasting, in a direction subparallel to the strike of the targeted ore-bearing reef. These tabular geometries are common in the deep gold mines of the Witwatersrand Basin (sedimentary rocks) and intermediate-depth platinum mines of the Bushveld Complex (igneous rocks) in South Africa (see Figure S1 in the Supporting Information for the photograph of a typical tabular platinum stope). The stope height is typically about 1–1.5 meters, much smaller than the span, which may extend for several hundred meters or more. Thus, a stope excavation is often viewed as a thin slit embedded within an elastic or inelastic rock mass (Cook, 1976; Jooste et al., 2023; McGarr, 1971a, 1971b), analogous to a Griffith-type crack in brittle solids (Jaeger et al., 2007). High stress concentrations near the edges of an excavation in the deep stopes could give rise to the formation of subvertical fractures, typically extending vertically for several tens of meters and spaced at less than one meter, which develop roughly parallel to the advancing stope faces (Figure 1b) (Cook, 1976). Such a fracture zone forms within both the hanging wall above and the footwall below the stope (Cook, 1976), and can extend several meters ahead of the stope face as well (Malan, 1999). As the mining face advances, dislocation movements along these fracture planes may cause differential block motions in the hanging wall and footwall, which are further manifested as closure or even contact between the stope roof and floor in regions away from the face (Malan, 1999; McGarr, 1971a).

In this paper, the term "closure" refers to relative movement of the hanging wall and footwall. In the intermediate-depth platinum mines, the rock mass behaviour of the igneous rocks of the Bushveld Complex is dominated by multiple sets of pre-existing joints. In this case, as the mining face advances, discontinuous movements occur along the joints in the hanging wall which can lead to large panel collapses.

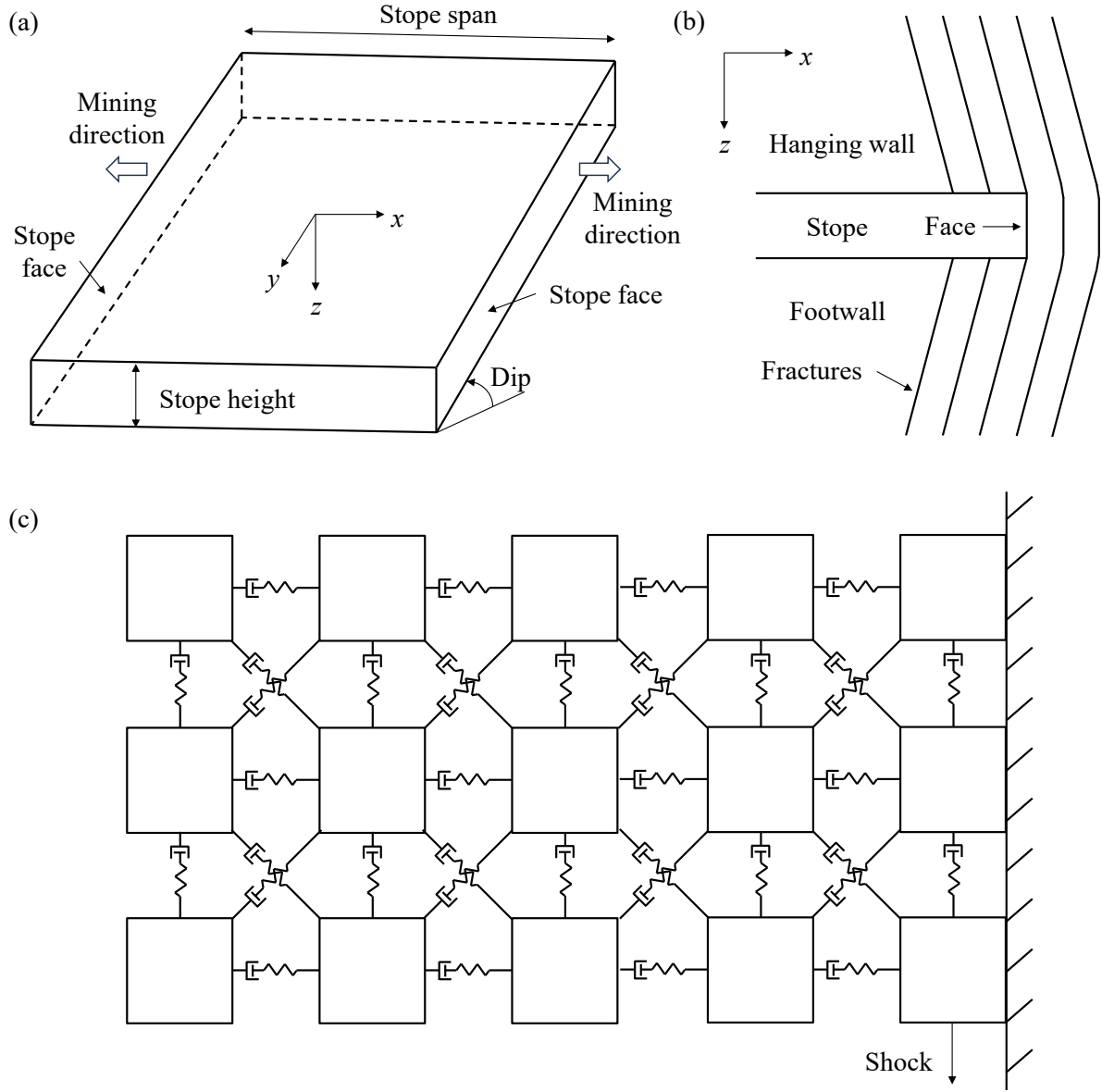

**Figure 1.** (a) Schematic of a typical parallel-sided dipping tabular stope. (b) Cross-sectional view of the stope, showing subvertical fractures generated near the edges. (c) Conceptual picture of the rock mass around the stope as a complex system composed of numerous blocks interacting via cohesive or frictional contacts, with block interactions governed by a time-dependent response function (note that the diagram is for illustrative purposes only; blocks do not necessarily have equal size and the frictional interface on the right side does not have to be vertical).

We conceptualize the rock mass surrounding a stope excavation as a complex system composed of numerous rock blocks interacting through cohesive and frictional contacts (Figure 1c). This conceptual picture mirrors the classical slider block model, a simplified framework commonly used in the earthquake community to conceptualize the interplay between heterogeneity and interactions, leading to cascading events reminiscent of domino chains or avalanches (Burridge & Knopoff, 1967; Rundle et al., 2003). It is also reminiscent of the classical block theory, which is well accepted in the geomechanics community for rock mass stability analysis (Goodman, 1995). It additionally resonates with dislocation theory, which models stope deformation as emerging from the collective interactions of dislocations within rock masses. This perspective has been effectively employed to study both gradual and abrupt deformations in mine stopes (McGarr, 1971a, 1971b). The number of blocks in the system is assumed to be large, which is justified by the small spacing between subvertical fracture planes relative to the stope span, as well as the presence of reef-parallel bedding planes, in the gold mines and by the many intersecting joints in the platinum mines. Building upon this assumption, we establish a mean-field theoretical framework grounded in the principles of statistical physics and derive analytical solutions to approximately describe the macroscopic behaviour of the complex rock block system in response to mining and related activities.

2.2 Epidemic-type triggering process

The stope closure activity arises from a combination of external forces, such as blasting and excavation, and internal effects, where each previously moved block can trigger movements of adjacent blocks through stress redistribution. This influence of a block on its nearby blocks is often not instantaneous, due to the time-dependent nature of the relevant geomechanical processes like subcritical crack growth (Brantut et al., 2013) and rate- and state-dependent friction (Marone, 1998). This latency can be described by a "bare" memory kernel $\phi(t-\tau)$, giving the probability that the movement of a block at time $\tau$ leads to the movement at a later time $t$ of another block in direct interaction with the first moved block. This bare memory kernel $\phi(t-\tau)$ encapsulates the fundamental macroscopic response time governing how long it takes for a block to be triggered into motion due to its interaction with a previously activated neighbor. In other words, it describes the distribution of waiting times between cause and action for a block to move. Based on many empirical observations such as Andrade's law of material creep (Andrade, 1910) and Omori's law of aftershock activity (Utsu et al., 1995), as well as various theoretical derivations based on the constitutive laws of subcritical crack growth (Shaw, 1993), rate- and state-dependent friction (Dieterich, 1994), and material rheology (Nechad et al., 2005a), we assume that $\phi(t-\tau)$ takes the form of a power law that typically characterizes long-memory processes (Sornette & Helmstetter, 2003):

$$\phi(t-\tau) = \frac{\theta c^\theta}{(t-\tau)^{1+\theta}}, \text{ with } 0 < \theta < 1 \text{ and for } t-\tau > c. \qquad (1)$$

The exponent $\theta$ governs the decay rate of past influences, effectively setting the memory span of the interaction kernel. The constant $c$ is a small characteristic time scale defining the onset of the power law decay and reflecting the time required for rupture (Kagan & Knopoff, 1981) due to brittle creep (Perfettini & Avouac, 2004), viscous deformation (Freed & Lin, 2001), pore pressure diffusion (Lindman et al., 2006), and/or frictional slip (Dieterich, 1994). Here, the lower bound of $\theta$ ensures that $\phi(t-\tau)$ is normalisable, i.e., $\int_{\tau+c}^{+\infty} \phi(t-\tau)dt = 1$, while the upper bound is based on documented empirical data (Nechad et al., 2005a; Hirata, 1987) revealing that $\theta$ rarely exceeds unity.

Let us now consider the block triggering process within the rock mass surrounding a stope excavation. Starting from an initial moved block, referred to as the "mother" block (also called a "key block" in the South African mining industry), which first moves due to either external forces (e.g., blasting) or internal fluctuations, it may trigger the movements of its first-generation neighbor "daughter" blocks, which subsequently trigger their own daughter blocks to move, continuing the cascade. We employ a mean-field approximation that represents block-to-block triggering interactions through an epidemic-type framework, modelled as a conditional self-excited point process (Hawkes & Oakes, 1974). This process has an exact mapping onto a branching process and can be mathematically described in terms of its average dynamics as (Sornette, 2006b; Sornette & Helmstetter, 2003):

$$v(t) = V(t) + n\int_{-\infty}^{t} \phi(t-\tau)v(\tau)d\tau = V(t) + n\int_{-\infty}^{t} \Phi(t-\tau)V(\tau)d\tau. \qquad (2)$$

Here, $v(t)$ is the average closure rate (also called velocity hereafter) evolving over time $t$; $V(t)$ is the exogenous activation that is not triggered by any epidemic effect within the system; $\Phi(t-\tau)$ is the "dressed" or renormalized memory kernel, which encodes the effective influence of a block set exogenously in motion at time $\tau$ on subsequent triggering at time $t$. This kernel integrates the multi-step dynamics from the exogenous input $V(\tau)$ to the endogenous response $v(t)$, and accounts for all possible generations of triggered block motions within the resulting triggering cascade; $n \geq 0$ is the branching ratio defined as the average number of first-generation triggered daughter block motions per mother block motion. Equation (2) is the equation for the first-order moment (or average) of the velocity, whose underlying dynamics is governed by a self-excited point process (Hawkes & Oakes, 1974). The first equality in equation (2) highlights that the current velocity $v(t)$ is influenced by all prior block motions, mediated by the bare memory kernel $\phi(t-\tau)$ that governs the direct triggering within each individual mother-daughter block pair, resulting in a self-consistent integral equation. The second equality in equation (2) presents the formal solution to this integral equation since the velocity does not appear anymore in the integral term. It describes how $v(t)$ arises from all past exogenous activations $V(\tau)$ with $\tau < t$, mediated by the dressed memory kernel $\Phi(t-\tau)$, collectively incorporating all generations of block interaction cascades.

The branching ratio $n$ depends on the network topology of the block system and the spreading behaviour of disturbances within the rock mass, thereby reflecting the system's maturation. For $n < 1$, the system is in the subcritical regime (Harris, 1963; Sornette, 2006a), where each block that moves induces on average fewer than one subsequent block to move in direct lineage. Consequently, the number of triggered blocks eventually decays to zero in the absence of external forcing. In this subcritical regime, the energy released by moving blocks is smaller than the energy required to trigger them. For $n > 1$, the system is in the supercritical regime (Harris, 1963; Sornette, 2006a), where the number of triggered masses on average grows exponentially with time (Helmstetter & Sornette, 2002) or even faster (Sornette & Helmstetter, 2002). In this regime, the energy released by the movement of a block typically surpasses the energy needed to initiate its movement. For $n = 1$, the system is in the critical regime, sitting at the borderline between the subcritical and supercritical regimes. In this state, the number of blocks triggered by a single block motion is distributed according to a heavy-tailed power law distribution (Saichev et al., 2005), reflecting a very large susceptibility to external shocks. In this paper, we focus on the subcritical and critical regimes with $n \leq 1$ to ensure stationarity under steady-state external forcing, whereas the supercritical regime $n > 1$ related to the occurrence of a catastrophic failure (Helmstetter & Sornette, 2002; Lei & Sornette, 2024; Sornette & Helmstetter, 2002) will be briefly explored in Section 4.

2.2 Solutions of the mean-field equation of epidemic-type triggering process

We present the solutions of the mean-field equation (2) to characterize the macroscopic behaviour of rock masses in response to external or internal shocks. First, let the rock mass be subjected to a strong external shock, e.g., a blasting event, where the exogenous activation $V(t)$ is represented by a Dirac function:

$$V(t) = V_c \delta(t - t_c). \tag{3}$$

Here, $V_c$ is the amplitude of the impulse occurring at time $t_c$ and $\delta$ is a unit impulse. Substituting equation (3) into the second integral of equation (2) yields:

$$v(t) = V_c \delta(t - t_c) + n \int_{-\infty}^{t} V_c \delta(\tau - t_c) \Phi(t - \tau) d\tau = n V_c \Phi(t - t_c), \text{ for } t > t_c, \tag{4}$$

which indicates that the post-peak dynamics are fully governed by the dressed memory kernel $\Phi(t-\tau)$ that can be derived as the solution of the Green's function of the first equality of equation (1) (Sornette & Helmstetter, 2003):

$$\Phi(t - t_c) = V_c \delta(t - t_c) + n \int_{-\infty}^{t} V_c \phi(t - \tau + t_c) \Phi(\tau - t_c) d\tau. \tag{5}$$

The solution is obtained by applying the Laplace transform to equation (5), which provides the Laplace transform of $\Phi(t-\tau)$. Taking the inverse Laplace transform then yields (Helmstetter & Sornette, 2002):

$$v(t) \propto \Phi(\tau - t_c) \propto \begin{cases} 1/(t-t_c)^{1-\theta}, & \text{for } c < t - t_c < t^* \\ 1/(t-t_c)^{1+\theta}, & \text{for } t - t_c > t^* \end{cases}. \tag{6}$$

where $t^*$ is a characteristic crossover time given by (Helmstetter & Sornette, 2002):

$$t^* = c \left[ \frac{n \Gamma(1-\theta)}{|1-n|} \right]^{1/\theta} \propto |1-n|^{-1/\theta}, \tag{7}$$

where $\Gamma(\cdot)$ is the gamma function. One can see that, as $n \to 1$ (in the critical regime), $t^* \to +\infty$, meaning that the early-time response ($t-t_c < t^*$) dominates; instead, if $0 < n < 1$ (in the subcritical regime), $t^*$ has a finite value, with the system characterized by a co-existence of both early-time response ($t-t_c < t^*$, where triggering cascades thrive) and late-time response ($t-t_c > t^*$, where triggering cascades diminish).

Without a significant external event, a peak may also spontaneously arise in the velocity evolution, from the combined effect of continuous stochastic fluctuations driven by minor noisy external perturbations and an amplified sequence of epidemic cascades driven by internal interactions. The average velocity trajectory before and after the endogenous peak $v_c$, conditioned on the existence of this peak at time $t_c$, is expressed by (Sornette & Helmstetter, 2003):

$$v(t | v(t_c) = v_c) \approx \frac{v_c}{\text{var}(v_c)} \text{cov}(v(t), v_c) = \frac{v_c}{\text{var}(v_c)} \int_{-\infty}^{\min(t, t_c)} \Phi(t - \tau) \Phi(t_c - \tau) d\tau, \tag{8}$$

for both $t < t_c$ and $t > t_c$. Considering equation (6), we can finally obtain:

$$v(t) \propto 1/(t-t_c)^{1-2\theta}, \text{ for } c < |t-t_c| < t^* \text{ (or equivalently for } n \to 1). \tag{9}$$

For $n < 1$ (subcritical regime), the system's response is primarily driven by random fluctuations, behaving essentially as a noise process, and can be described by:

$$v(t) \propto 1/|t-t_c|^{\theta}, \text{ for } |t-t_c| > t^*. \tag{10}$$

2.3 Endo-exo classification

Based on the solutions derived above, the stope closure velocities near a peak at time $t_c$ can be represented by a generalized power law finite-time singularity function:

$$v(t) \propto 1/|t-t_c|^p, \qquad (11)$$

where the exponent $p$ depends on the parameter $\theta$ and the regime delineated by the characteristic crossover time $t^*$. This framework enables classification of velocity peaks into four distinct types (Figure 2), distinguished by the origin of disturbance (endogenous versus exogenous) and the level of criticality (subcritical versus critical):

(1) Exogenous-subcritical peak ($n < 1$ and $t-t_c > t^*$), characterized by $p = 1+\theta$. In this case, the cascading propensity is limited ($n < 1$), and the exogenously induced velocity jump at time $t_c$ does not cascade beyond the first few generations of triggered blocks (Fig. 2a).

(2) Exogenous-critical peak ($n \approx 1$ and $c < t-t_c < t^*$), characterized by $p = 1-\theta$. In this case, the system is in a critical state ($n \approx 1$), such that the exogenously induced velocity jump at time $t_c$ cascades throughout the block system, where activated blocks trigger further movements of their neighbors, and so forth (Fig. 2b).

(3) Endogenous-subcritical peak ($n < 1$ and $|t-t_c| > t^*$), characterized by $p = 0$. The closure activity arises not from an external shock but from endogenous interactions. Limited cascading occurs ($n < 1$), resulting in a velocity peak that lacks apparent precursory or recovery signatures (Fig. 2c).

(4) Endogenous-critical peak ($n \approx 1$ and $c < |t-t_c| < t^*$), characterized by $p = 1-2\theta$. Here, the closure activity arises from endogenous growth and interaction within a critical system ($n \approx 1$), leading to movement cascades that exhibit an approximately symmetrical power law acceleration and deceleration patterns around the peak (Fig. 2d).

This classification of system responses into four qualitatively distinct regimes emerges from the interplay of the long-memory process as described by equation (1) and the mean-field epidemic cascade throughout the system as captured by equation (2). It can be seen that the recovery following an endogenous-critical peak (with a smaller exponent $p = 1-2\theta$) is slower than that following an exogenous-critical peak (with a larger exponent $p = 1-\theta$). This more persistent influence of an endogenous-critical peak results from its longer precursory preparation, which permeates the system more deeply than an exogenous-critical peak. Note that exogenous peaks do not exhibit apparent precursors (see Figures 2a and 2b), as they are triggered by external events that impose sudden shocks to the system. Based on this taxonomy, velocity peaks can be readily classified based on their distinct precursory and recovery patterns. Specifically, exogenous peaks exhibit power law relaxation without apparent precursory acceleration, with exogenous-critical peaks having $p < 1$ and exogenous-subcritical ones having $p > 1$; on the other hand, endogenous critical peaks are characterized by symmetrical power law acceleration and deceleration with $p < 1$, while endogenous subcritical peaks exhibit no apparent precursory and recovery dynamics with $p$ being around 0. Thus, the distinct characteristics of each peak type enable differentiation between endogenous versus exogenous origins of episodic rock creep, based on the system's dynamical response around peaks rather than potentially ambiguous correlations with external events. It is important to clarify that the differentiation here pertains to the initial triggering source (whether internal or external), rather than the drive of the system's response, which is always governed by endogenous self-excited triggering processes within the rock mass (see more discussions in Section 4.2).

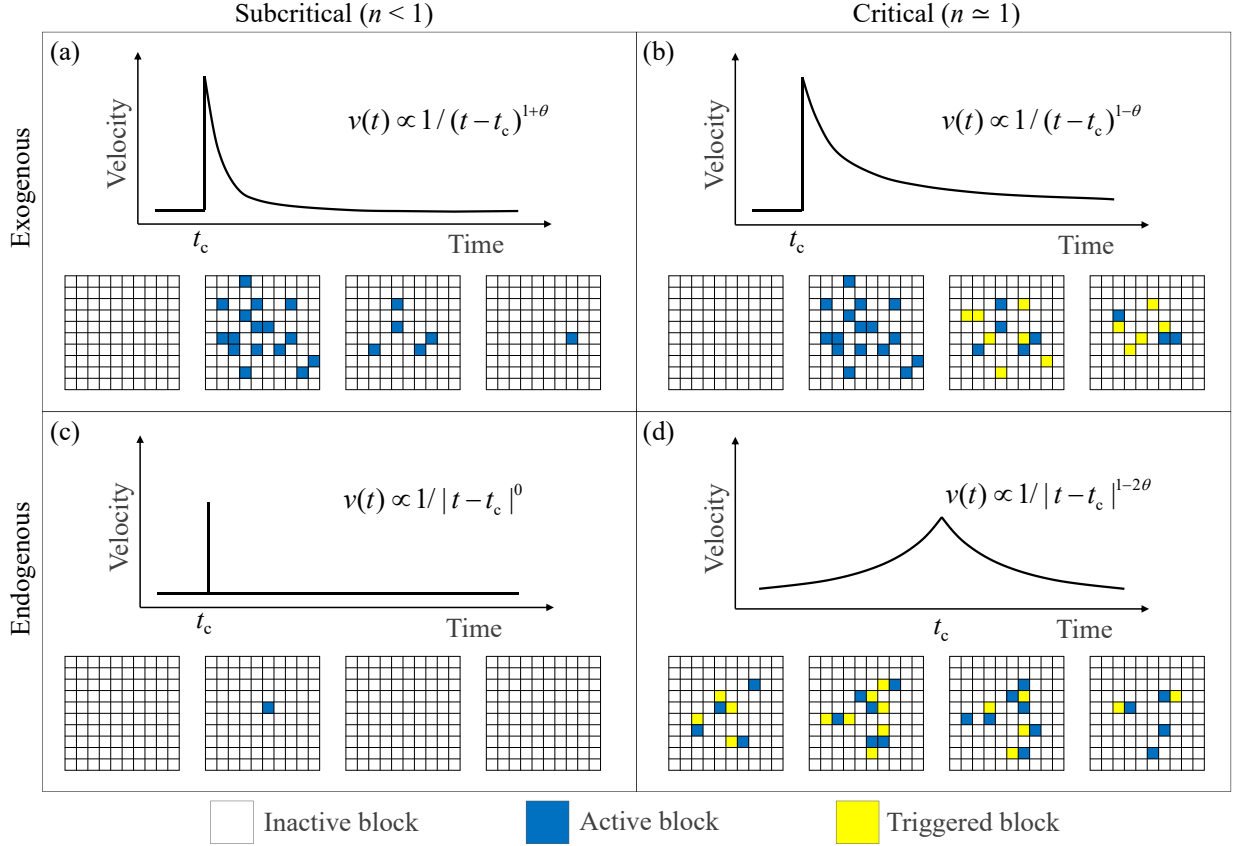

**Figure 2.** Classification of episodic creep dynamics based on the origin of disturbance (endogenous versus exogenous) and the level of criticality (subcritical versus critical). There are four distinct types of power law velocity dynamics $v(t)$ around a peak at time $t_c$ and they are all related to a single parameter $\theta$. The series of snapshots below each velocity trajectory illustrates the evolution of triggering processes within a rock block system (with inactive blocks, active blocks, and triggered blocks color-coded) underpinning the velocity time history.

2.4 Data acquisition and analysis approach

We conduct a comprehensive case study to test the endo-exo theory using stope closure measurements from a platinum mine in South Africa, where multiple panels mined within the Upper Group 2 (UG2) reef at a depth of approximately 1200 m below the ground surface (Malan et al., 2007). The UG2 reef is a chromitite layer within the Bushveld Complex, one of the world's largest layered mafic intrusions, renowned for its high concentrations of platinum group metals including platinum, palladium, and rhodium. The UG2 panels were mined beneath the Merensky Reef (see Figure S2 in the Supporting Information for the layout of UG2 panels), another platinum-rich layer within the Bushveld Complex that had been previously exploited in this region. As UG2 mining approached and extended beneath the Merensky remnant, significant closure occurred within the extending stope, which was closely monitored at multiple panels using clockwork closure meters instrumented between the hanging wall and footwall (Malan et al., 2007). Such closure meters can continuously record stope closure, providing measurements as a function of time at a sampling frequency significantly greater than that of mining face advance cycles (Malan, 1998; Malan et al., 2007; Malan & Napier, 2021) (see Figure S3 in the Supporting Information for the field setup and operational mechanisms of the closure meters instrumented).

We derive velocity time series $v(t)$ by differentiating the closure time series with respect to time. Subsequently, we fit the velocity data around a peak to the following equation:

$$v(t) = \frac{C}{(t-t_c)^p} + v_r, \qquad (12)$$

where $C$ is a constant and $v_r$ is the residual velocity when the system has fully recovered from external perturbations. The determination of this residual velocity for a stope during mining is subject to uncertainties, because the rock mass has very rare opportunities to completely recover from an external perturbation (e.g., blasting operations, excavation activities, and seismic events) before the next event occurs. Here, we estimate the residual velocity by first detecting troughs in the velocity time series and then define the residual velocity associated with a given peak as the minimum of the two nearest troughs (with one before the peak and one after the peak). This residual velocity is found typically around $3\times10^{-5}$ mm/min in our analysis, which is compatible with past field observations of near-zero closure rate during the December holidays when blasting and excavation cease in the mine. To estimate $C$ and $p$, we use the least square method to minimize the sum of squared errors:

$$s = \sum_{t_i} \varepsilon(t_i)^2 = \sum_{t_i} \left\{ \ln\left[v(t_i) - v_r\right] - \ln C + p \ln |t_i - t_c| \right\}^2, \qquad (13)$$

where $v(t_i)$ is the empirically measured velocity at time $t_i$. We then set the partial derivatives $\partial s/\partial(\ln C)$ and $\partial s/\partial p$ to be both equal to zero, leading to solve a linear system of two equations with the two unknowns $C$ and $p$. Thus, through the least squares method, we can fit equation (12) to the time series of net velocity $v(t)-v_r$ around each peak to estimate the associated power law exponent $p$, with its standard deviation derived from the confidence interval of the fit. Unless stated otherwise, velocity in the following sections refers to the net velocity.

## 3 Results

Figures 3a and 4a present the closure measurements for UG2 Panel 1N (where "1" denotes the panel number and "N" indicates the northern sector of mining), mined toward the Merensky remnant (see Figure S4 in the Supporting Information for the location of Panel 1N with respect to this remnant), from June to August 2005. During this period, the face advanced 23.1 m in total, such that part of the UG2 face reached beneath the boundary of the Merensky remnant. The data shown in Figures 3a and 4a were recorded at two different locations, as the closure meter was relocated closer to the advancing face on August 3, 2005. The stope displayed a step-like deformation pattern over time, marked by multiple episodes of temporary acceleration and deceleration (Figures 3a and 4a). Notably, significant jumps in closure occurred immediately after each blasting event, followed by a gradual deceleration sustained over time. Post-peak velocity relaxation following these blast-induced exogenous-critical peaks obeys a power law, with an exponent of $p = 0.54\pm0.03$ at the first measurement position (Figure 3b) and $p = 0.57\pm0.20$ at the second position (Figure 4b). These $p$ values generally show a good agreement, but a more pronounced dispersion is observed in the data collected at the second position, likely due to its closer proximity to the Merensky remnant and the resulting stronger stress redistribution and/or more heterogeneous ground response. This may also account for the significantly higher closure rate at the second measurement position, where almost 120 mm of closure was reached over 17 days (Figure 4a), compared to around 140 mm over 40 days at the first position (Figure 3a). The panel remained stable despite such high rates of closure (see Figure S5 in the Supporting Information for a field photo of this panel).

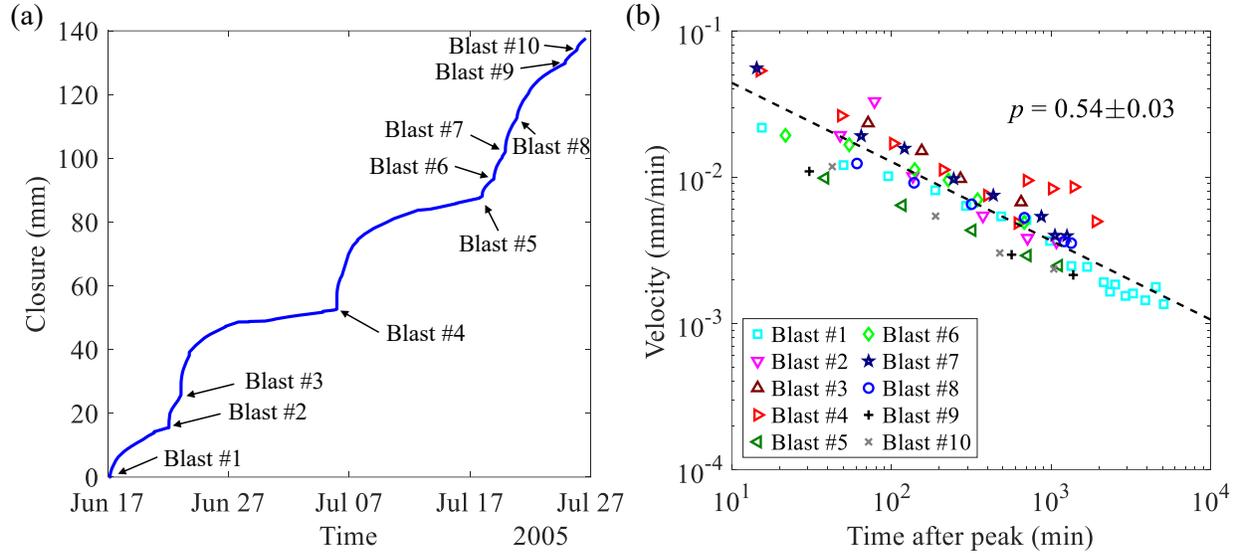

**Figure 3.** (a) Closure measurements at the first instrumentation position for UG2 Panel 1N during the period from June 17 to July 27, 2005. The start and end of the time series respectively correspond to distances of 9.4 m and 20.0 m from the advancing stope face. (b) Post-peak velocity relaxation following blast-induced exogeneous-critical peaks, with the dashed line indicating the power law fit.

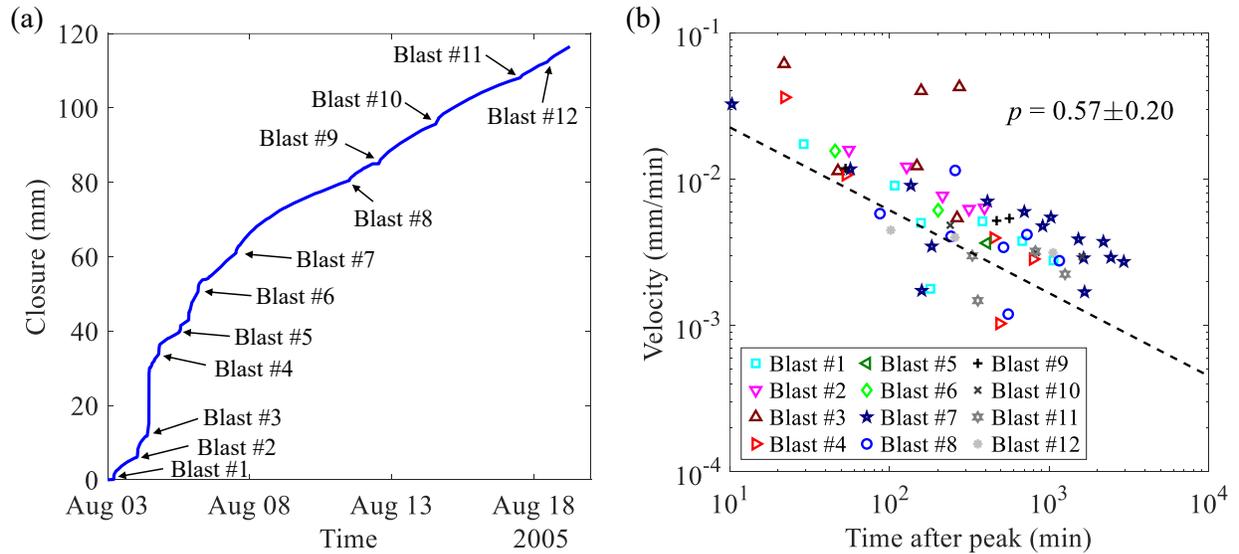

**Figure 4.** (a) Closure measurements at the second instrumentation position for UG2 Panel 1N during the period from August 03 to August 19, 2005. The start and end of the time series respectively correspond to distances of 11.8 m and 24.3 m from the advancing stope face. (b) Post-peak velocity relaxation following blast-induced exogeneous-critical peaks, with the dashed line indicating the power law fit.

We further analyze the closure data collected for UG2 Panel 4S (where "4" denotes the panel number and "S" indicates the southern sector of mining) over a 35-day period from September 21 to October 26, 2005 (Figure 5). The panel was already situated below the Merensky remnant when the monitoring began (see Figure S6 for the panel location). During the monitoring period, the face advanced 14.1 m in total. The stope also exhibited a step-like deformation pattern over time, characterized by a series of acceleration-deceleration episodes associated with successive blasting operations (Figure 5a).

Post-peak velocity relaxation following these blast-induced exogenous-critical peaks obeys a power law, with an exponent of $p = 0.69\pm0.07$ (Figure 5b). In addition to those blast-induced acceleration-deceleration sequences, an acceleration crisis spontaneously emerged on October 17, 2005, in the absence of any blasting event (Figure 5c). This coincided with a catastrophic failure event in the adjacent Panel S5 (see Figure S7 in the Supporting Information for a photo of this collapse) and was followed by a gradual deceleration over the next few days (Figure 5c). This endogenous-critical peak appears to be preceded by a progressively accelerating power law growth of velocity and succeeded by an approximately symmetrical power law decline of velocity, which are characterized in general by a common exponent of $p = 0.36\pm0.06$ (Figure 5d). Note that only two data points are available for the pre-peak stage, while earlier velocities show no correlation with the peak, possibly because this endogenous-critical event originated in the adjacent Panel S5 and had not yet influenced Panel 4S at that time.

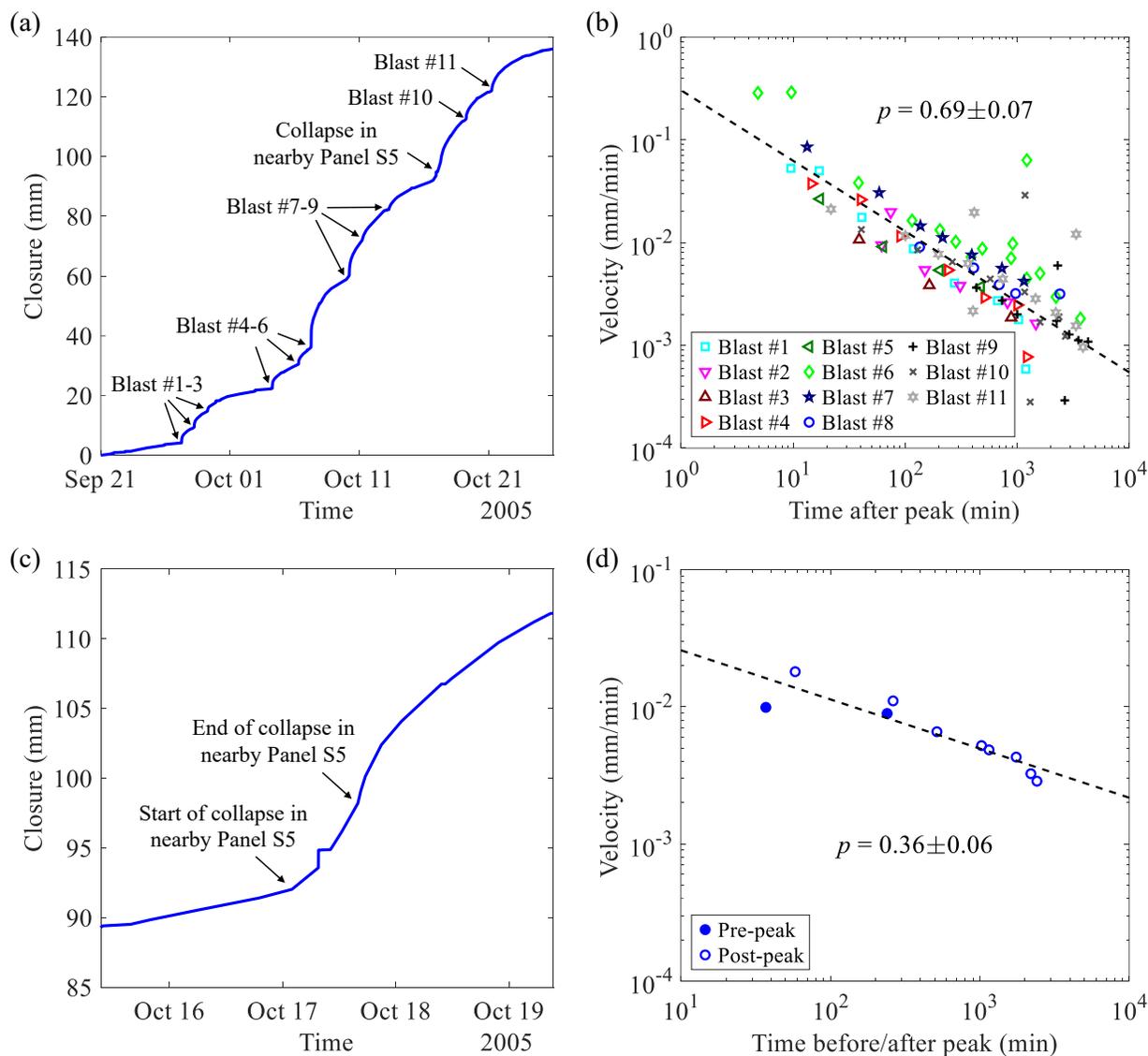

**Figure 5.** (a) Closure measurements for UG2 Panel 4S during the period from September 21 to October 26, 2005. The start and end of the time series respectively correspond to distances of 7.4 m and 21.5 m from the advancing stope face. (b) Velocity relaxation following blast-induced exogeneous-critical peaks, with the dashed line indicating the power law fit. (c) Local view of the closure curve around an endogenous

acceleration crisis during which the adjacent Panel S5 experienced a collapse. (d) Pre-peak acceleration and post-peak relaxation of velocity around an endogenous-critical peak, with the dashed line indicating the power law fit.

Figure 6a shows the closure behaviour recorded at the first measurement position in UG2 Panel 6N for the period from January 24 to February 5, 2007, during which the face advanced 7.8 m. The post-peak velocity relaxation following blast-induced exogenous peaks also conforms to a power law, with an exponent of $p = 0.56\pm0.07$ (Figure 6b). An acceleration crisis developed on February 01, 2007, during which an abrupt jump in closure was observed (Figure 6c). The corresponding velocity peak is surrounded by an essentially noisy stationary velocity trajectory with no apparent precursory and recovery trends, with the power law exponent $p$ close to 0 (Figure 6d), thus manifesting as an endogenous-subcritical peak.

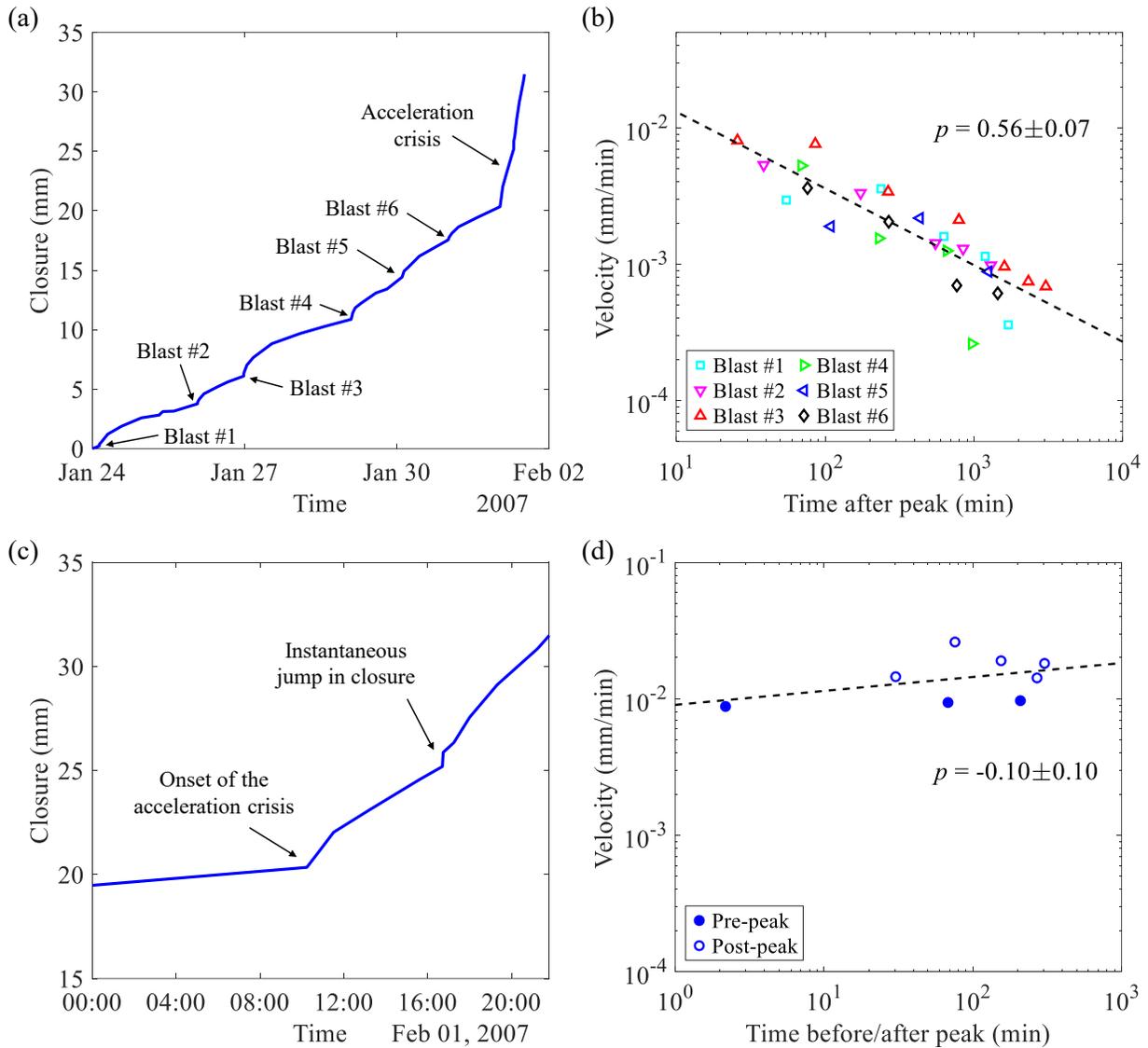

**Figure 6.** (a) Closure measurements at the first instrumentation position for UG2 Panel 6N during the period from January 24 to February 05, 2007. The start and end of the time series respectively correspond to distances of 17.8 m and 25.6 m from the advancing stope face. (b) Velocity relaxation following blast-induced exogeneous-critical peaks, with the dashed line indicating the power law fit. (c) Local view of the
*Lei et al.* Preprint at arXiv     13*Lei et al.* Preprint at arXiv     13

closure curve during an acceleration crisis. (d) Pre-peak and post-peak velocity dynamics around an endogenous-subcritical peak, with the dashed line indicating the power law fit.

Figure 7a further presents the closure data collected at the second measurement position in UG2 Panel 6N from February 9 to February 11, 2007. During this short period, the face advanced 2 m before the panel was abandoned due to the high closure rate encountered as it reached beneath the Merensky remnant. Interestingly, the velocity relaxation following blast #1 exhibits a two-branch power law scaling behavior: the early-time response within approximately 100 mins is characterized by an exponent $p = 0.76\pm0.36$, while the late-time response is associated with $p = 1.35\pm0.11$ (Figure 7b). These characteristics indicate that it is an exogenous-subcritical peak. On the other hand, the velocity recovery following blast #2 exhibits a single power law trend with an exponent $p = 0.66\pm0.05$ (Figure 7b), indicating that it is an exogenous-critical peak.

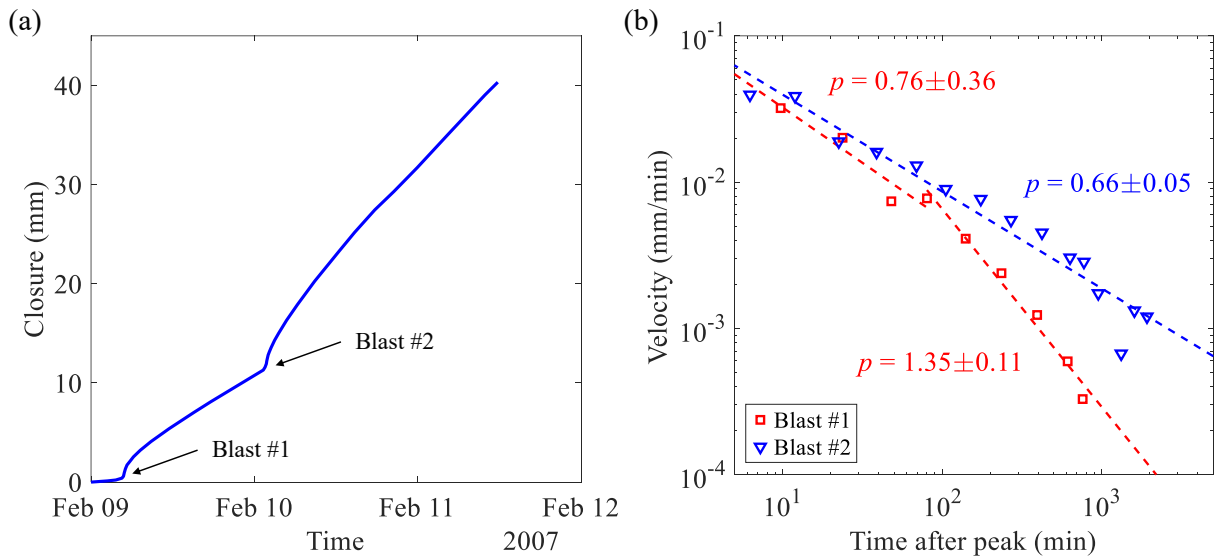

**Figure 7.** (a) Closure measurements at the second instrumentation position for UG2 Panel 6N during the period from January 09 to February 11, 2007. The start and end of the time series respectively correspond to distances of 19.6 m and 21.6 m from the advancing stope face. (b) Velocity relaxation following blast-induced exogeneous-critical/subcritical peaks, with the dashed line indicating the power law fit.

Thus, all the four types of episodic rock creep, namely endogenous/exogenous-subcritical/critical, have been found in the closure measurement data. Interpreting these results in light of equations (6), (9), and (10), the various exponents $p$ in expression (12) can be mapped onto $1+\theta$, $1-\theta$, $0$, or $1-2\theta$ and can thus be associated with the four distinct types of velocity peaks. Remarkably, all these exponents emerge naturally from the endo-exo framework and depend solely on a single parameter $\theta \approx 0.35\pm0.1$. This provides a strong validation for our endo-exo theoretical framework. Although some velocity peaks, particularly the endogenous ones, are derived from limited data and some individual fits may lack high reliability, the combined analysis across multiple closure data sets yields a coherent picture, and a robust and well-constrained estimate of the $\theta$ parameter.

## 4 Discussion

### 4.1 Mechanisms behind the observed parameter $\theta$

We have proposed an endo-exo framework for the analysis of episodic rock creep in deep underground mines. It offers a quantitative classification of episodic rock creep into four fundamental types, each characterized by distinct precursory and recovery signatures, yet unified by a single common parameter $\theta$ which governs the bare memory kernel describing first-generation triggering of block motions through direct block-to-block interactions. All four power law regimes of episodic dynamics were identified in stope closure measurement data from a platinum mine in South Africa, with $\theta$ found to be around 0.35. This $\theta$ value differs from the prediction by the classical fiber bundle model which assumes load sharing and long-range stress transfer mechanism, yielding $\theta = 0$ (Nechad et al., 2005a, 2005b; Saichev & Sornette, 2005). It is also smaller than the value $\theta = 0.5$ predicted by the random stress model that assumes a normal diffusion of stress fluctuations (Kagan & Knopoff, 1987; Lei & Sornette, 2024). Here, we extend this random stress model to account for anomalous stress diffusion, thereby accounting for the observed $\theta$ value in the current study.

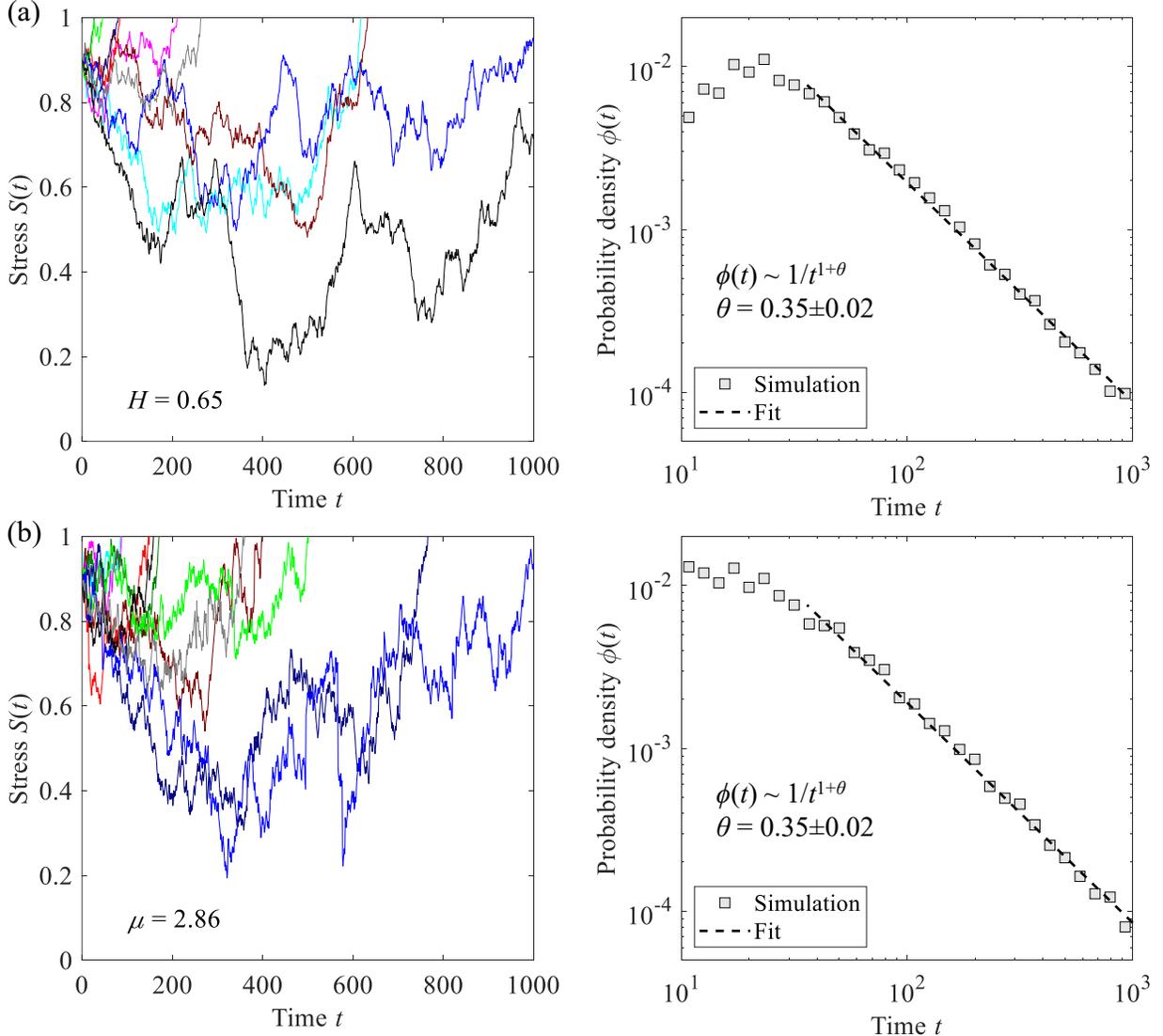

**Figure 8.** Random walk simulations of stress fluctuations with increments $s$ (a) obeying a fractional Brownian motion with a Hurst exponent of $H = 0.65$ and a characteristic increment amplitude of $s^* = 0.01$, or (b) drawn from a Pareto distribution with a shape parameter of $\mu = 2.86$ and a minimum increment size of $s_{min} = 0.01$. In the simulation, stress $S(t)$ evolves as a random walk starting from an initial value $S_0 = 0.9$ with each step taking a unit time, and the process stops until $S(t)$ for the first time exceeds the critical threshold $S_c = 1.0$. Left panels give the stress trajectories of 10 realizations which are arbitrarily selected from 10000 simulated realizations. Right panels show the probability density function of waiting times $\phi(t) \propto 1/t^{1+\theta}$, derived from the random walk simulations, with $\theta = 0.35$ obtained for both models.

Without loss of generality, let us adopt a dimensionless form and consider that, at time $t = 0$, a mother block moves and perturbs its neighboring first-generation daughter blocks, which are initially subject to a stress level $S_0$ marginally below the critical threshold $S_c$ required to trigger movement. Suppose that the subsequent stress at each first-generation daughter block motion fluctuates due to random increments $s$ originating from various possible sources (e.g., microcracking, stress redistribution, microseismic tremors), which may be described by a fractional Brownian random walk (Mandelbrot & Van Ness, 1968) associated with a Hurst exponent $0 < H < 1$ and a characteristic increment amplitude $s^*$. For $H = 0.5$, stress fluctuations follow normal diffusion, whereas $H \neq 0.5$ leads to anomalous diffusion with positively correlated increments for $H > 0.5$ (persistent behavior) and negatively correlated increments for $H < 0.5$ (anti-persistent behavior). Then, the waiting time for a daughter block to start moving is determined by the first-passage time of its stress $S(t)$ exceeding the critical threshold $S_c$. The corresponding probability density function of this first-passage time asymptotically follows a power law $\phi(t) \propto 1/t^{(2-H)}$ (Rangarajan & Ding, 2000; Rosso & Zoia, 2014). Comparing this with equation (1) gives $\theta = 1-H$. This relationship is confirmed by random walk simulations with parameters $S_0 = 0.9$, $S_c = 1$, $s^* = 0.01$, and $H = 0.65$, which yield $\theta = 0.35 \pm 0.02$ (Figure 8a). Physically, the value of $\theta = 0.35$ reflects persistent (positively correlated) stress fluctuations, where stress increments follow a non-Markovian process and are more likely to continue in the same direction, either towards the critical threshold or away from it. This behaviour captures a memory effect in stress evolution, driven by processes such as microcrack coalescence, localized stress redistributions, and coherent microseismic perturbations. Consequently, the waiting times for daughter blocks to move are broadly distributed, with intermittent bursts of correlated stress accumulation driving block movements.

We consider another random walk model whose increments $s$ now follow a Markovian process but are drawn from a Pareto distribution $\mathbb{P}(s) = \mu s_{min}^\mu / s^{\mu+1}$, where $\mu$ is the shape parameter (power law exponent) and $s_{min}$ is the minimum increment size. For $0 < \mu < 2$, the increments of the random walk have an infinite variance and the first passage is dominated by rare large jumps that grow larger and larger over time, scaling as $s_{max} \sim s_{min} t^{1/\mu}$. As a result, the first-passage time follows the Lévy regime scaling $\phi(t) \propto 1/t^{1+1/\mu}$ (Chechkin et al., 2006). A comparison with equation (1) then gives $\theta = 1/\mu$. For $\mu > 2$, the increments of the random walk have finite mean and variance, so the leading scaling properties of the random walk converge to those of standard Brownian motion, in accordance with the central limit theorem (Sornette, 2006a). Consequently, the first-passage time distribution scales asymptotically as $\phi(t) \propto 1/t^{3/2}$ with $\theta = 0.5$ to leading order (Redner, 2001). Nevertheless, due to the fat-tailed nature of the Pareto distribution governing stress increments, convergence to the scaling laws of the standard Brownian motion is markedly slow. This is because rare large jumps continue to significantly influence stress evolution at intermediate times, before the central limit theorem fully takes effect (Sornette, 2006a, chapter 4). This gives rise to a quasi-Lévy regime in the first-passage time distribution at intermediate times. We anticipate the crossover time from this intermediate-time quasi-Lévy regime to the long-time Brownian regime to be approximately $t_{cross} \approx [(S_c-S_0)/s_{min}]^\mu$, based on the criterion that the typical maximum jump $s_{max}$ must be sufficiently large to span the threshold distance $S_c$–$S_0$. It is important to note that the crossover time $t_{cross}$ should not be confused with the characteristic crossover time $t^*$ defined in equation (7), which delineates the transition between early-time and late-time responses in the exogenous-subcritical regime. Random walk simulations using $S_0 = 0.9$, $S_c = 1$, $\mu = 2.86$, and $s_{min} = 0.01$ (yielding an estimated $t_{cross} \approx 724$) confirm the existence of this intermediate quasi-Lévy regime. A power law fit to the waiting time distribution up to $t = 1000$ yields $\theta = 1/\mu = 0.35\pm0.02$ (Figure 8b). This result suggests that the observed $\theta = 0.35$ in the studied platinum mine may also arise from the intermittent and bursty nature of stress evolution within rock masses, driven by Lévy-type sudden and significant stress redistributions associated with microcracking or microslip events, in addition to or rather than from regular Brownian-type super-diffusive processes.

It is worth noting that the above random walk models also provide a natural interpretation for the parameter $c$ in the bare memory kernel as defined by equation (1). Specifically, parameter $c$ represents the characteristic time scale over which random walkers bridge the gap between the initial stress level and the critical threshold. Accordingly, a smaller gap corresponds to a lower $c$ value.

### 4.2 Further insights into endo-exo interactions

Our results indicate that most large velocity peaks in this platinum mine dataset are exogenous-critical, suggesting that the rock mass around the mining stope is often in a critical state. This is consistent with previous conjectures and observations that rock masses in deep tabular mines are subject to high stress concentrations, particularly in the vicinity of the stope face, and are often critically stressed (Cook, 1976; Lucier et al., 2009; McGarr, 2000). Under such critical conditions, the rock mass response to strong external blasting events is primarily driven by cascades of triggered block movements where each triggered block motion in turn initiates the movement of subsequent blocks across multiple generations of triggering. Consequently, the collective response of the rock mass is slower and more sustained, characterized by a power law decay with a small exponent $1-\theta$, in contrast to the faster relaxation of individual blocks, which follows a power law decay with the larger exponent $1+\theta$.

We have also identified an exogenous-subcritical creep event (Figure 7b) that initially undergoes approximately 100 mins of slower exogenous-critical relaxation characterized by the smaller exponent $1-\theta$ before transitioning into a faster exogenous-subcritical phase marked by the larger exponent $1+\theta$. Substituting this characteristic time $t^* \approx 100$ mins together with $\theta \approx 0.35$ and an estimate of $c \approx 1$-10 mins into equation (7), we obtain $n \approx 0.62$-$0.78$. Here, our estimate of $c$ is based on an inspection of the power law decay trends around peaks in the data, and it falls within the typical range of minutes to days reported in the literature (Helmstetter & Sornette, 2002; Utsu et al., 1995). This comparatively low branching ratio $n$ indicates that, although the rock mass around a mining stope is often critically stressed, it may not always remain at criticality. Similar observations have been reported recently for other systems, such as earthquake faults (Nandan et al., 2021) and landslides (Lei & Sornette, 2024). This is in contradiction with the conventional concept of self-organized criticality (Bak, 1996), which postulates that the crustal rock is permanently evolving at a critical state and all the events are generated by the same underlying process, making the prediction of large events impossible. More sophisticated models of self-organized criticality, that integrate earthquake dynamics with fault network evolution, have overturned this view, revealing that the stress field typically remains far from its rupture threshold (Cowie et al., 1993; Sornette et al., 1994). Our results confirm the existence of different regimes, with characteristic signatures emerging during transitions that may signal the approach of catastrophic events, an insight with significant potential for improving predictive capabilities. Notably, the exogenous-subcritical peak at Panel 6N was preceded by multiple exogenous-critical peaks (Figure 6) and then followed by an exogenous-critical peak (Figure 7), after which the stope was abandoned due to hazardous conditions associated with high closure rates (Malan et al., 2007). This reflects the inherently intermittent and fluctuating nature of the deep mine system, where transitions between critical and subcritical states likely arise from local heterogeneities, stress redistributions, and transient external perturbations that temporarily reduce the system's susceptibility, driving it into a brief subcritical phase. Subsequently, stress re-accumulation and cascading block movements restore critical conditions. This resonates with previous field observations reporting a slight arrest in the rate of closure or tilt before major events (Cook, 1976; McGarr & Green, 1975). Such a dynamical, non-stationary behavior reveals the system's intrinsic complexity and highlights the necessity of continuous monitoring to provide valuable early-warning signals for hazard mitigation, as also emphasized in other studies (Malan et al., 2007; Malan & Napier, 2021).

In addition, the dataset reveals the presence of endogenous-critical and endogenous-subcritical peaks (Figures 5b and 6d), although they are much less prominent than the exogenous peaks, suggesting that these endogenous events are either rare or overshadowed by the more dramatic exogenous events. Notably, the endogenous-critical peak at UG2 Panel 4S coincided with the collapse in the adjacent Panel 5S, implying a strong endogenous coupling of the rock masses across the panels. Interestingly, similar creep episodes, lacking any apparent exogenous triggers yet exhibiting clear acceleration-deceleration trends, were previously also observed in the tiltmeter measurements at a deep gold mine in South Africa (McGarr & Green, 1975). These episodes, typically lasting several days, were termed "anomalous" tilt due to the limited understanding of their underlying mechanisms. Within our endo-exo framework, they can be rationalized as manifestations of endogenous-critical creep dynamics.

Our results confirm that, for a system in the subcritical/critical regime, it is possible to differentiate between endogenous and exogenous origins of system responses. This differentiation relies on the fact that a complex system responds differently to exogenous shocks compared to endogenous ones, as demonstrated by our theoretical analysis of Section 2.2, schematically illustrated in Figure 2, and further supported by our application to the comprehensive dataset acquired at the platinum mine. The validity of this differentiation for complex systems has been extensively demonstrated across much broader contexts, including geological, social, economic, and biological systems (Crane & Sornette, 2008; Helmstetter et al., 2003; Helmstetter & Sornette, 2002; Lei & Sornette, 2024; Roehner et al., 2004; Sornette et al., 2003, 2004, 2009). To avoid ambiguity, we emphasize that the classification into endogenous and exogenous origins pertains solely to the initial trigger, whether internally generated or externally imposed, and not to the overall evolution of the system. These systems are in fact subject to continuous external driving through boundary stresses or bulk perturbations, while their internal state evolves endogenously via mechanisms such as dislocation motion, microcracking, damage accumulation, healing, and thermo-hydro-mechanical-chemical processes.

It is crucial to point out that endogenous and exogenous factors are always linked, as long as the system is not driven exclusively by external factors (i.e., $n \neq 0$) and remains coupled to external influences, such as a continuous flow of noise fluctuations and occasional significant external disturbances. Let us first imagine an extreme scenario in which the system is entirely driven by external inputs ($n = 0$). In this case, external events would only cause isolated responses within the system, leaving the rest of it entirely inactive and devoid of any fluctuations. Of course, such an idealized system does not exist in reality. Whenever $n > 0$, which generally reflects an internally organized system with coupled constituent components, a large external event tends to provoke additional events within the system, as observed in both the exogenous-subcritical and exogenous-critical regimes. For instance, in the exogenous-critical regime, the initial shock originates externally, but it can trigger a potentially large cascade of subsequent events, reflecting the system's subsequent endogenous dynamics. On the other hand, in the endogenous-critical regime, the system is primarily driven by numerous small external stochastic fluctuations, which are then amplified into cascades of endogenously triggered events. These triggered events can lead to spontaneously growing activity that culminates in an endogenous peak. The activity dynamics before and after this endogenous peak are roughly symmetrical, in contrast to the pronounced asymmetrical pattern observed for an exogenous-critical peak. In both cases (given $n \leq 0$), the underlying physics and dynamics are fundamentally the same: the system's behavior is driven by self-excitation and internal triggering within the system. Our endo-exo classification framework explicitly recognizes the intrinsic coupling between exogenous events and endogenous triggering processes. This should not be confused with our effort to differentiate exogenous and endogenous origins, which specifically refers to the initial source that triggers the system's response, rather than the driving mechanism of these responses, which is consistently governed by self-excitation and triggering processes that are inherently endogenous.

### 4.3 Implications for forecasting violent rockbursts

Hitherto, our focus has been primarily on the "endo-exo" regime, where rock masses exhibit repeated episodes of acceleration and deceleration, driven by the interplay of exogenous perturbation and endogenous maturation. We now explore the last research question raised in Section 1: How does episodic rock creep relate to violent rockbursts?

As the stope excavation progresses, the surrounding rock masses may undergo a transition into the supercritical regime. This transition often arises due to endogenous maturation of the system (Sornette, 2006b), explaining why sometimes stope collapses occur in the absence of significant blasting or seismic activities (Malan et al., 2007). In this supercritical regime, the system dynamics typically follows the so-called Voight's equation (Voight, 1988, 1989):

$$\dot{v}(t) \propto v(t)^\alpha, \text{ with } \alpha > 1, \tag{14}$$

where the exponent $\alpha$ characterizes the degree of nonlinearity, and the condition $\alpha > 1$ ensures the presence of positive feedbacks, resulting in a finite-time singularity. This singular behavior can be seen by integrating equation (14), leading to (Lei & Sornette, 2025b):

$$v(t) \propto 1/(t_c - t)^p, \text{ with } p > 0, \tag{15}$$

which is in the same form as equation (11) (but only valid for $t < t_c$) with exponent $p = 1/(\alpha-1)$ and $t_c$ being the critical time of failure. A further integration of equation (15) gives the solution of the time-dependent stope closure $\Omega(t)$ (Lei & Sornette, 2025b):

$$\Omega(t) = A + B(t_c - t)^m, \text{ with } m < 1, \tag{16}$$

where $A$ and $B$ are constants, and $m = 1-p$ is the singularity exponent.

Such finite-time singular behavior arises in the mean-field epidemic-type formulation (2) when variations in the fertility of triggering blocks are taken into account. Mathematically, this corresponds to augmenting the bare memory kernel $\phi(t-\tau)$ defined in equation (1), with a fertility factor that quantifies the number of block motions triggered by a given initiating event. When fertilities follow a power law distribution with an exponent less than 1, the average branching ratio diverges. As shown by Sornette & Helmstetter (2002), this leads to the finite-time singular behavior described by expression (15).

We analyse the closure measurements from UG2 Panel 7N, where a catastrophic failure event occurred on 04 April 2007 (see Figure S8 in the Supporting Information for a field photo of this collapse) and the system's acceleration toward the final failure was well captured by the monitoring system (Figure 9a). We fit the pre-failure velocity time series to equation (15), obtaining $p = 0.91 \pm 0.22$ (Figure 9b), which is generally compatible with the theoretical prediction of $p = 1$ for $\theta > 0$ (Sornette & Helmstetter, 2002). Independently, we fit the pre-failure closure time series to equation (16), yielding $m = 0.12$ (Figure 9a, inset). Notably, these values approximately satisfy the expected relationship of $m = 1-p$. From them, we can further estimate $\alpha \approx 2.1$ (equation (14)), which aligns well with the typical $\alpha$ values of around 2 reported in the literature (Intrieri et al., 2019; Voight, 1988, 1989). This suggests that positive feedbacks indeed dominate the rock mass behaviour as it approaches the final failure. Such positive feedbacks may arise from crack growth in intact rocks (Kilburn, 2012; Main, 1999; Nemat-Nasser & Horii, 1982; Sammis & Sornette, 2002) and frictional instability along fault planes (Helmstetter et al., 2004; Noda & Chang, 2023), both of which have been documented in field observations at deep mines (Cook, 1976; McGarr, 1971b; McGarr et al., 1975, 1979; Ortlepp & Stacey, 1994; Spottiswoode & McGarr, 1975), with the former mechanism generating smaller events and the latter one producing relatively larger events (Richardson, 2002).

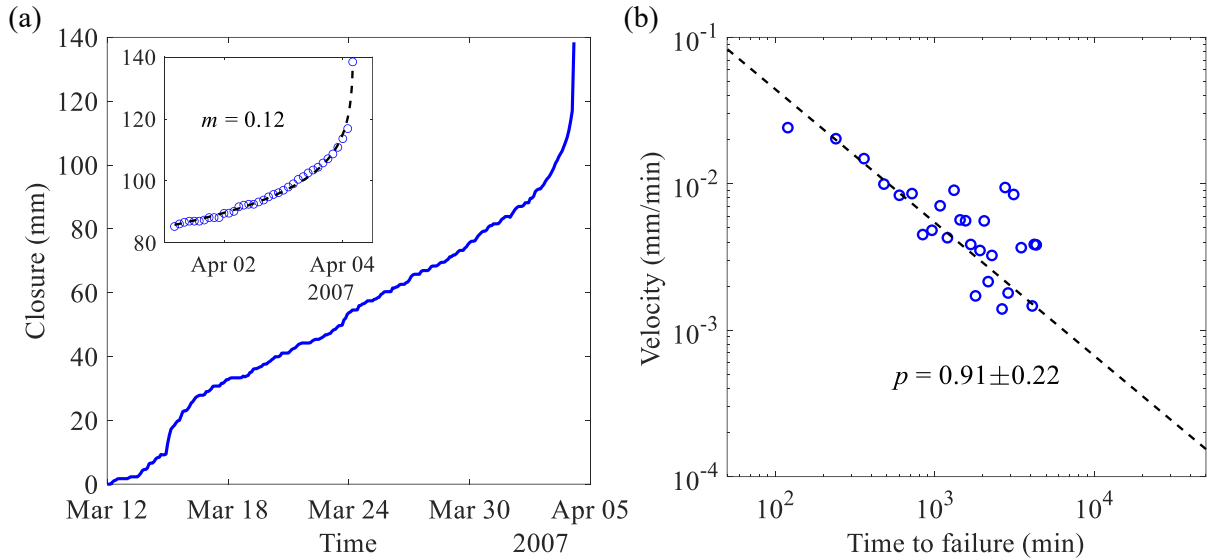

**Figure 9.** (a) Closure measurements for UG2 Panel 7N during the period from March 12 to April 04, 2007. The end of the time series corresponds the occurrence of a catastrophic collapse. Inset: closure evolution during the terminal phase prior to the final failure with the dashed line indicating the fit to a power law singularity model. (b) Pre-failure velocity as a function of time to failure (time flows from right to left), with the dashed line indicating the power law fit.

Different from the subcritical and critical regimes for which the endo-exo classification applies, when the system enters the supercritical regime ($n > 0$), an initial exogenous shock will cascade into a dominant endogenous dynamics, carrying a finite probability of triggering cascades of events whose number grows exponentially with time (Helmstetter & Sornette, 2002) or even faster (Sornette & Helmstetter, 2002). It is crucial to note that the underlying driving process remains the same as that in the subcritical and critical regimes, where the system's response is governed by self-excitation and triggering mechanisms. In this supercritical regime, different parts of the system are strongly coupled and interdependent (Sornette, 2002; Sornette & Ouillon, 2012), leading to the dominance of endogenous processes. In this regime, the system is characterized by high endogeneity, making it inherently fragile and sensitive to external perturbations. Even a minor external perturbation can trigger substantial amplification through endogenous positive feedback mechanisms, leading to catastrophic failure. It is worth noting that the dramatic response of the system to a shock does not mean that the exogenous factor itself becomes more significant; rather, it is the endogenous processes that grow increasingly important, amplifying the impact of an external event on the system and giving the illusory impression of a stronger exogenous influence.

The physical processes underlying the transition from the subcritical/critical regimes to the supercritical regime, which ultimately leads to catastrophic failures, may be related to damage accumulation under fatigue loading from repeated blasting operations and ongoing stope excavations. This endogenously driven failure can emit precursory signals, such as accelerated creep and log-periodic oscillations (Lei & Sornette, 2025b, 2025a; Ouillon & Sornette, 2000), which are highly informative for prediction, rendering violent rockbursts potentially predictable "dragon-kings" (a double metaphor for an event of a predominant impact like a "king" and born from a unique origin like a "dragon") (Lei et al., 2023; Sornette & Ouillon, 2012). However, violent rockbursts may occasionally appear as unpredictable "black-swans" (a metaphor for rare, high-impact events that come as a surprise) (Taleb, 2010). These black-swan events may result from unstable rupture along blast-generated inward-dipping fractures (i.e., inclined toward the stope face), where the stored strain energy cannot be sufficiently dissipated in advance (McGarr, 1971b). This contrasts with the more stable creep behavior along blast-generated outward-dipping fractures (i.e., inclined away from the stope face, as illustrated in Figure 1b) (McGarr, 1971a), where part of the stored strain energy can be gradually dissipated, often accompanied by identifiable precursory signals, prior to an eventual failure. It is also conjectured that black-swan rockbursts are more likely to occur in strong, brittle rock formations where limited deformation precedes failure, whereas dragon-king rockbursts tend to develop in more ductile settings that allow for progressive deformation and energy dissipation. This aligns with past extensive field observations in South African gold and platinum mines: in brittle, high-strength rocks such as Ventersdorp lavas and quartzites (with uniaxial compression strengths exceeding 400 MPa and 250 MPa, respectively), stope closure monitoring rarely revealed precursory acceleration before large seismic events. In contrast, in more ductile platinum mine settings (where the rock's uniaxial compression strength is typically below 150 MPa), sustained stable closure can be observed and may allow the detection of precursors. Black-swan scenarios pose greater challenges for monitoring and forecasting compared to endogenously driven dragon-king cases; however, acknowledging the limits of predictability is itself a valuable insight.

## 5 Conclusions

To conclude, we have developed a novel endo-exo theoretical framework to quantitatively diagnose episodic creep events in deep underground rock masses, capturing the interplay between external triggers (such as blasting and excavation) and internal processes (such as damage and healing). Our framework classifies episodic dynamics into four fundamental types, defined by the origin of disturbance (endogenous or exogenous) and the level of criticality (subcritical or critical), which are all governed by a single parameter $\theta$, which controls the four distinct temporal power law regimes surrounding a velocity peak. Our formulation effectively captures the fundamental endo-exo mechanisms while maintaining the flexibility needed for empirical calibration. Applied to the comprehensive monitoring data of a deep platinum mine in South Africa, this parameter $\theta$ was estimated to be around 0.35, which we interpret as reflecting persistent stress fluctuations arising from anomalous diffusion processes and bursty stress redistributions that drive episodic rock mass movements and interactions. Our parsimonious framework points at the existence of a deep quantitative relationship between episodic dynamics, external triggers, and internal processes in rock masses. This framework enables us to differentiate between endogenous and exogenous origins of rock creep responses, while also shedding light on their underlying physical mechanisms and their connection to catastrophic failures. Our findings provide critical insights into the fundamental mechanisms of episodic creep movements in rock masses and pave the way for enhanced rockburst hazard prediction and mitigation in deep mining environments. Our results also have broad implications for understanding similar episodic phenomena in other geophysical systems such as fault slow slip, landslide creep, volcanic unrest, and glacier surge.

**Acknowledgments**

Q.L. is grateful for the support from the Swedish Rock Engineering Research Foundation and State Key Laboratory of Intelligent Coal Mining and Strata Control. D.S. acknowledges partial support by the National Natural Science Foundation of China (Grant no. T2350710802, U2039202), and the Center for Computational Science and Engineering at Southern University of Science and Technology. Q.L. acknowledges the National Academic Infrastructure for Supercomputing in Sweden (NAISS), partially funded by the Swedish Research Council through grant agreement no. 2022-06725, for awarding this project access to the LUMI supercomputer, owned by the EuroHPC Joint Undertaking and hosted by CSC (Finland) and the LUMI consortium.

Supporting Information for

# Endo-exo classification of episodic rock creep in deep mines: Implications for forecasting catastrophic failure

Qinghua Lei[1], Daniel Francois Malan[2], Didier Sornette[3]

[1]Department of Earth Sciences, Uppsala University, Sweden
[2]Department of Mining Engineering, University of Pretoria, Hatfield, South Africa
[3]Institute of Risk Analysis, Prediction and Management, Academy for Advanced Interdisciplinary Studies, Southern University of Science and Technology, Shenzhen, China

**Contents of this file**

   Figures S1 to S8

**Introduction**

This document provides supporting information to complement the results and discussions in the main Letter. Figure S1 presents the photograph of a typical tabular platinum stope. Figure S2 shows the layout of UG2 panels mined toward and below a Merensky remnant. Figure S3 illustrates the field setup and operational mechanisms of the clockwork closure meters. Figure S4 shows the location of Panel 1N with respect to the Merensky remnant. Figure S5 shows the typical ground conditions in Panel 1N. Figure S6 shows the location of Panel 4S with respect to the Merensky remnant. Figure S7 presents a photo of Panel 5S which collapsed on 17 October 2005. Figure S8 presents a photo of Panel 7N which collapsed on 04 April 2007.

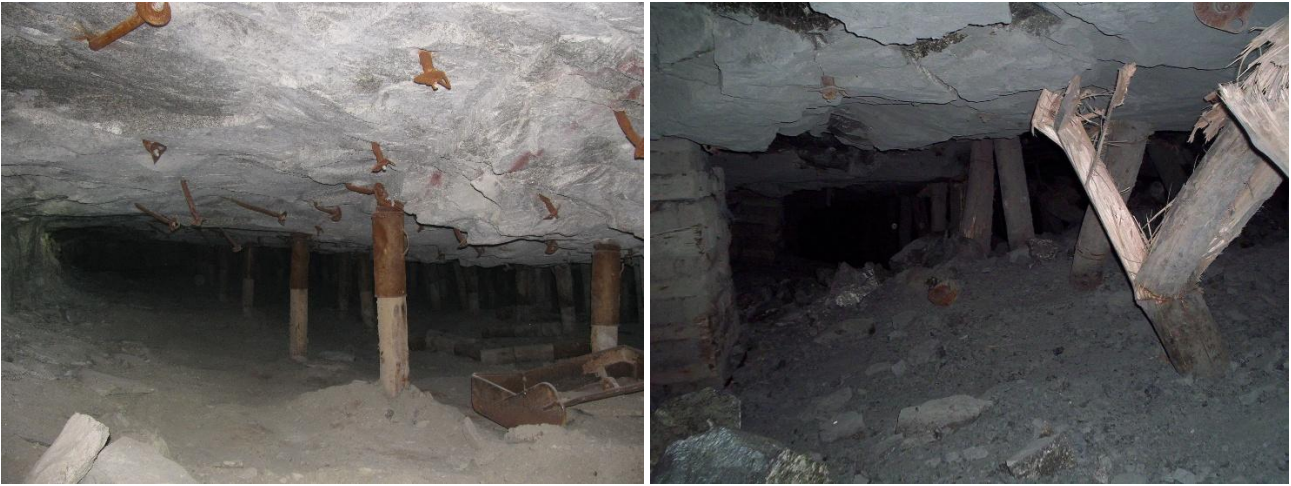

**Figure S1.** Photograph of a typical tabular platinum stope (left) and the hanging wall unravelling along the joints prior to a large panel collapse (right). There are many interacting blocks in the hanging wall. Note the evidence of significant closure and the broken elongates.

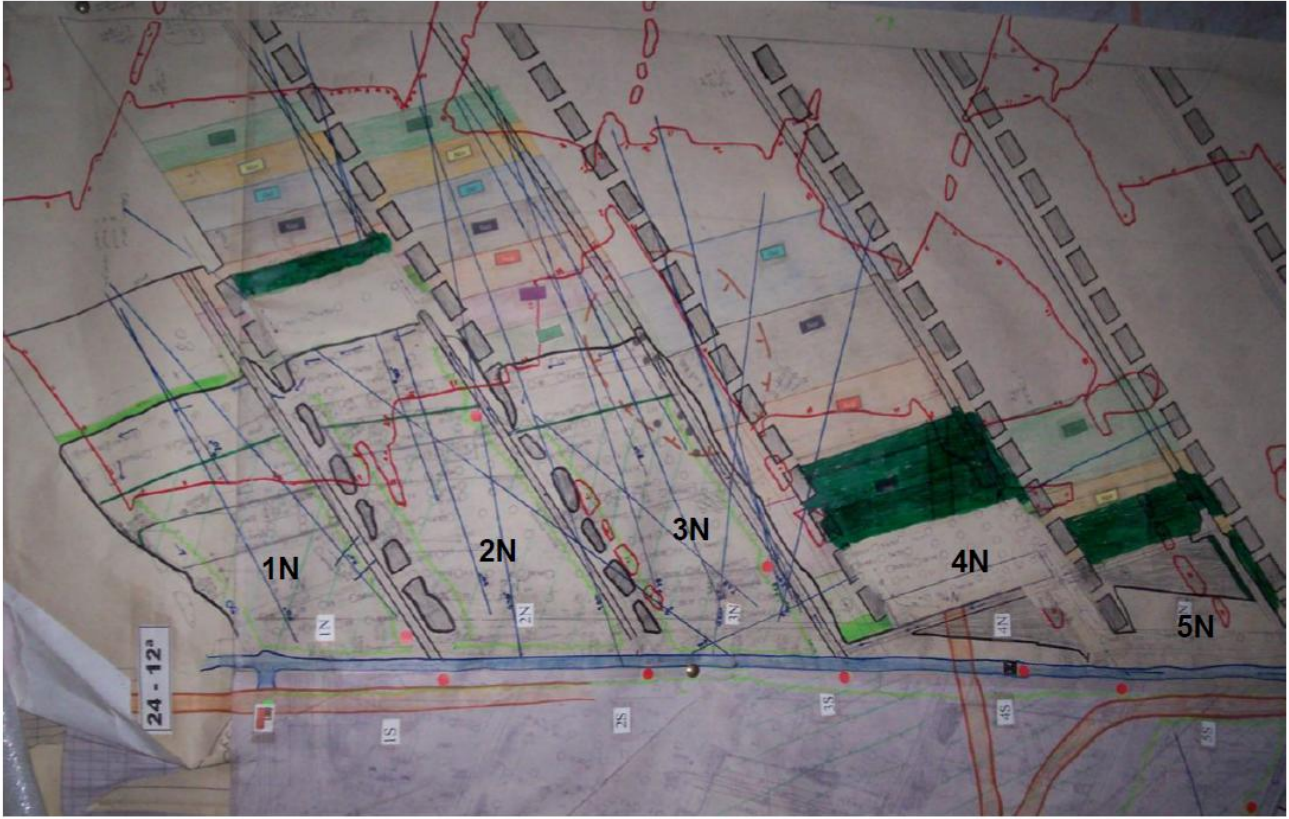

**Figure S2.** UG2 panels mined toward and below a Merensky remnant.

(a) (b)

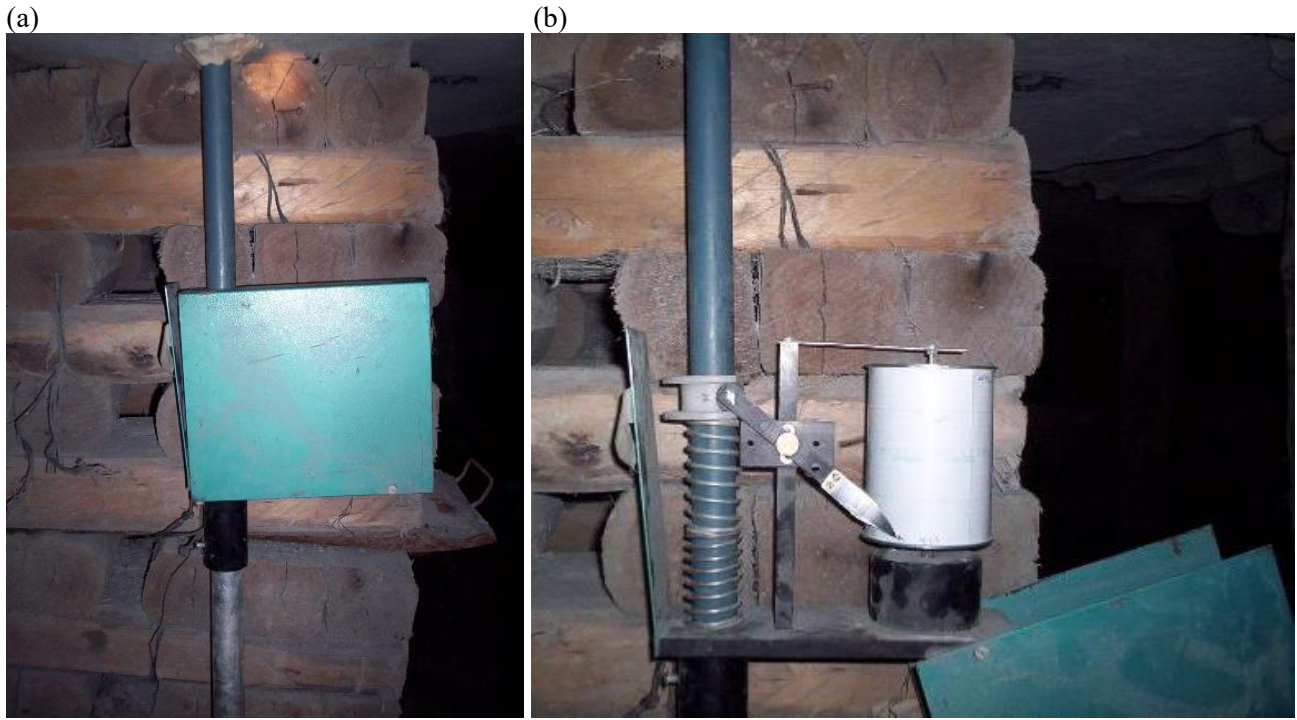

(c)

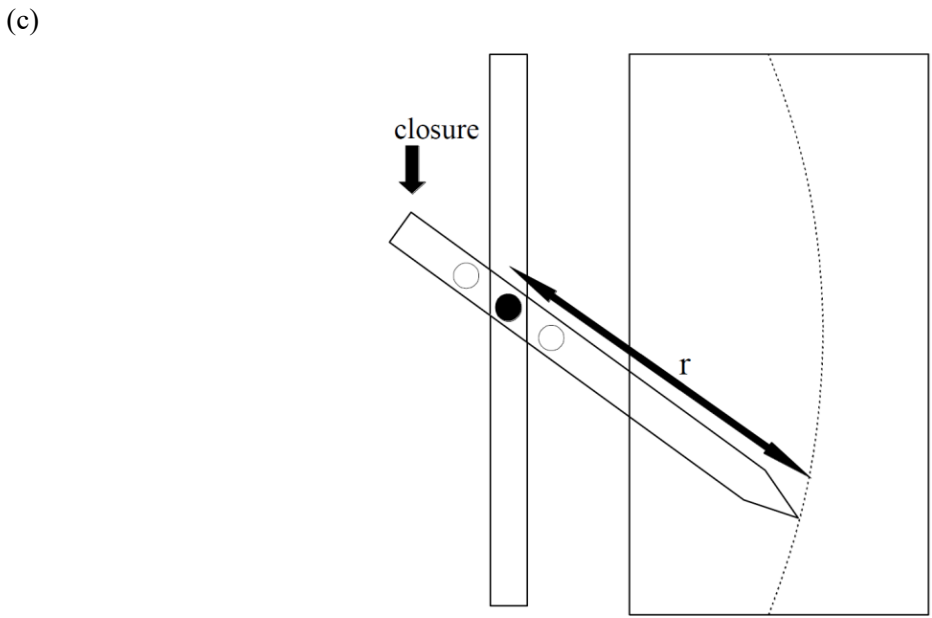

**Figure S3.** Photographs of a clockwork closure meter (a) in a closed box and (b) with the box open to reveal the mechanism, which is further illustrated in (c). It is composed of two pipes, one of which slides within the other under the resistance of a spring. The base of the instrument is attached to the lower pipe, while a guide affixed to the upper pipe moves the needle. The instrument is set up with the needle tip positioned near the lower edge of the drum's graph paper. The stope closure forces the upper pipe deeper into the lower pipe, causing the needle to move upwards on the paper. The needle arm features multiple pivot points, enabling adjustable amplification of the measured closure magnitude. The clock completes one full rotation over a certain period, producing a continuous record of the closure throughout that time. The raw data need to be corrected before use and more details of the correction method can be found in Malan (1998).

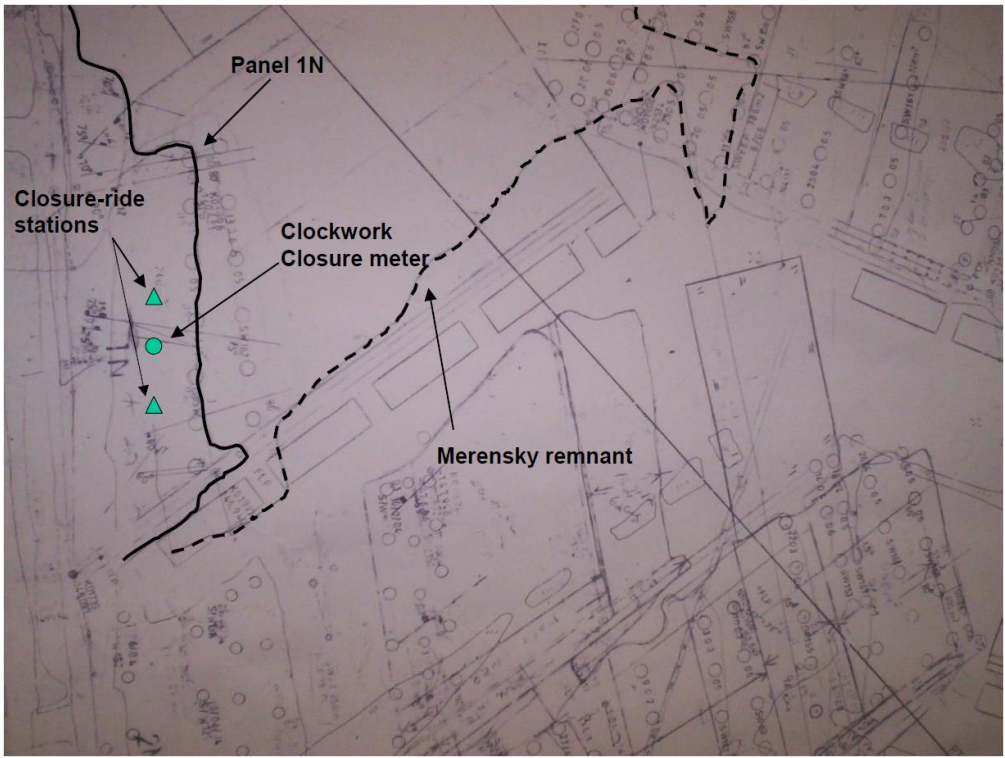

**Figure S4.** Location of Panel 1N with respect to the Merensky remnant.

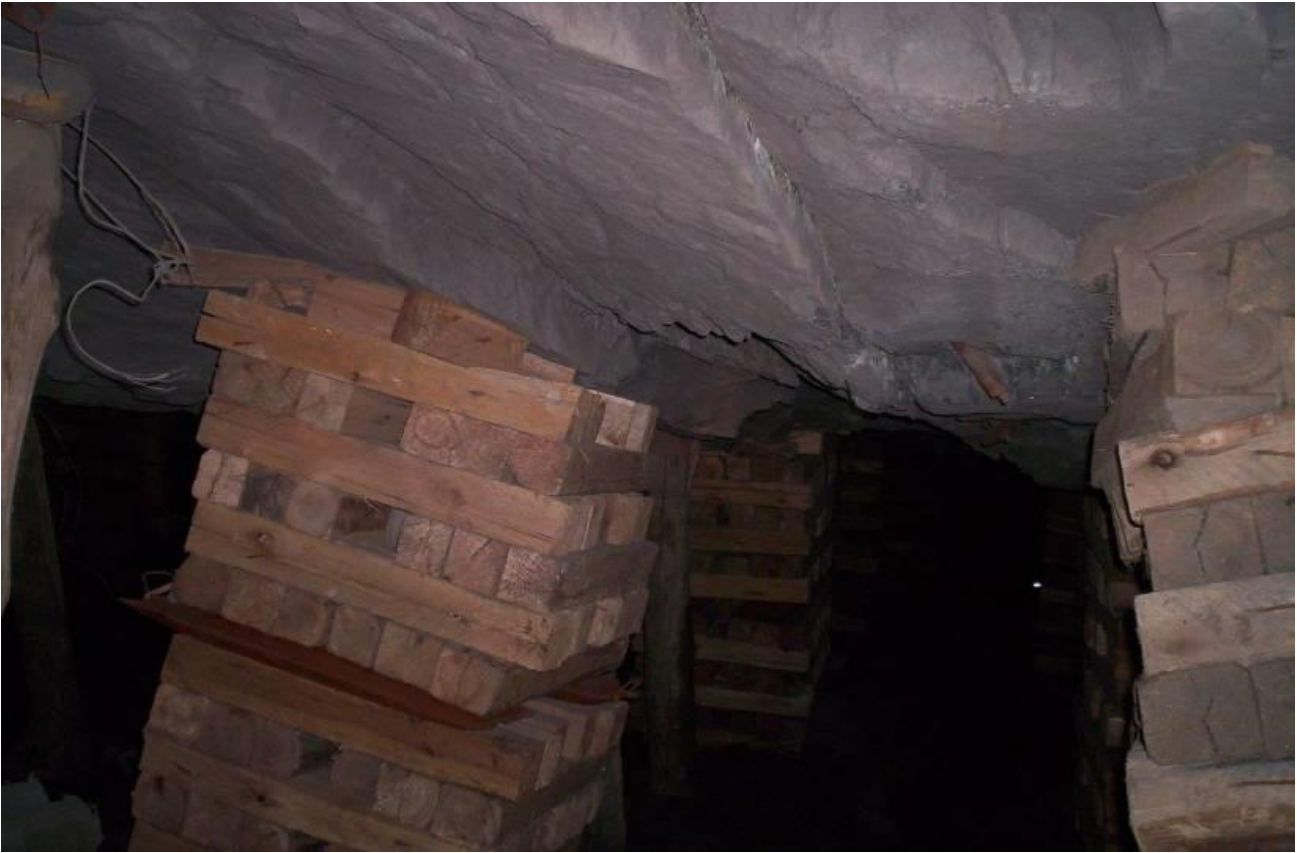

**Figure S5.** Typical ground conditions in Panel 1N.

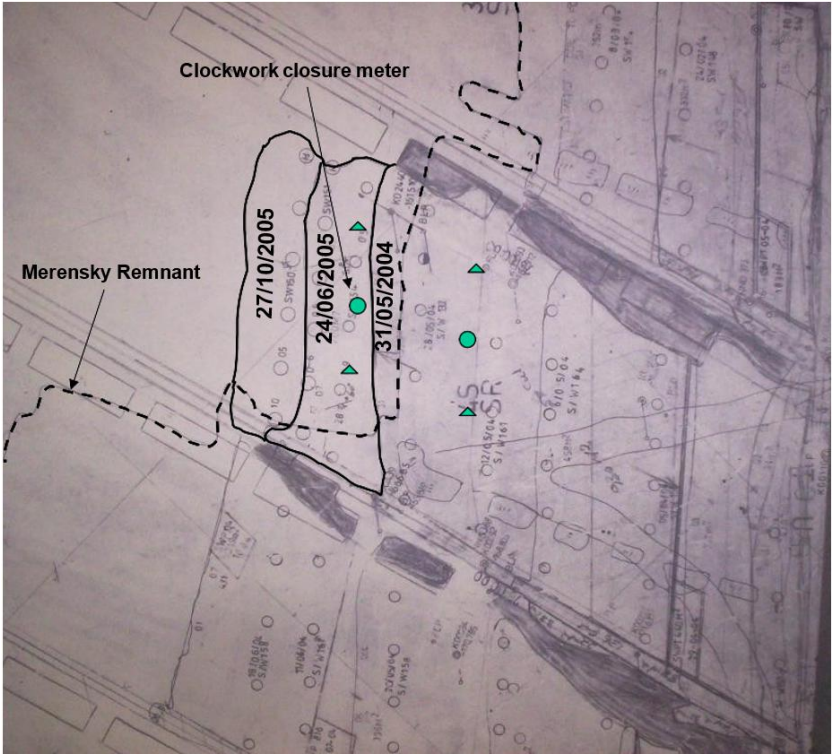

**Figure S6.** Location of Panel 4S with respect to the Merensky remnant.

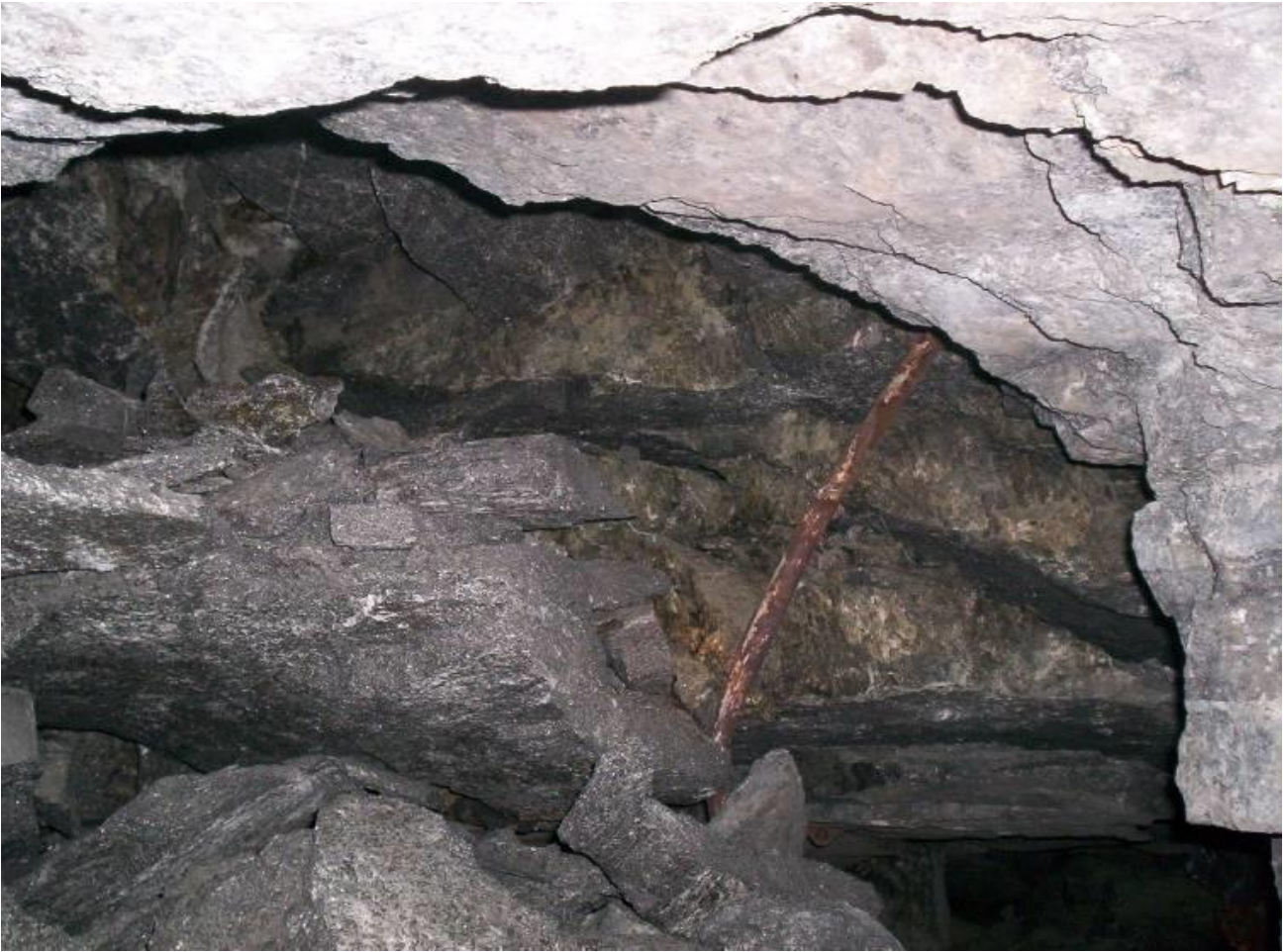

**Figure S7.** Collapse in Panel 5S on 17 October 2005.

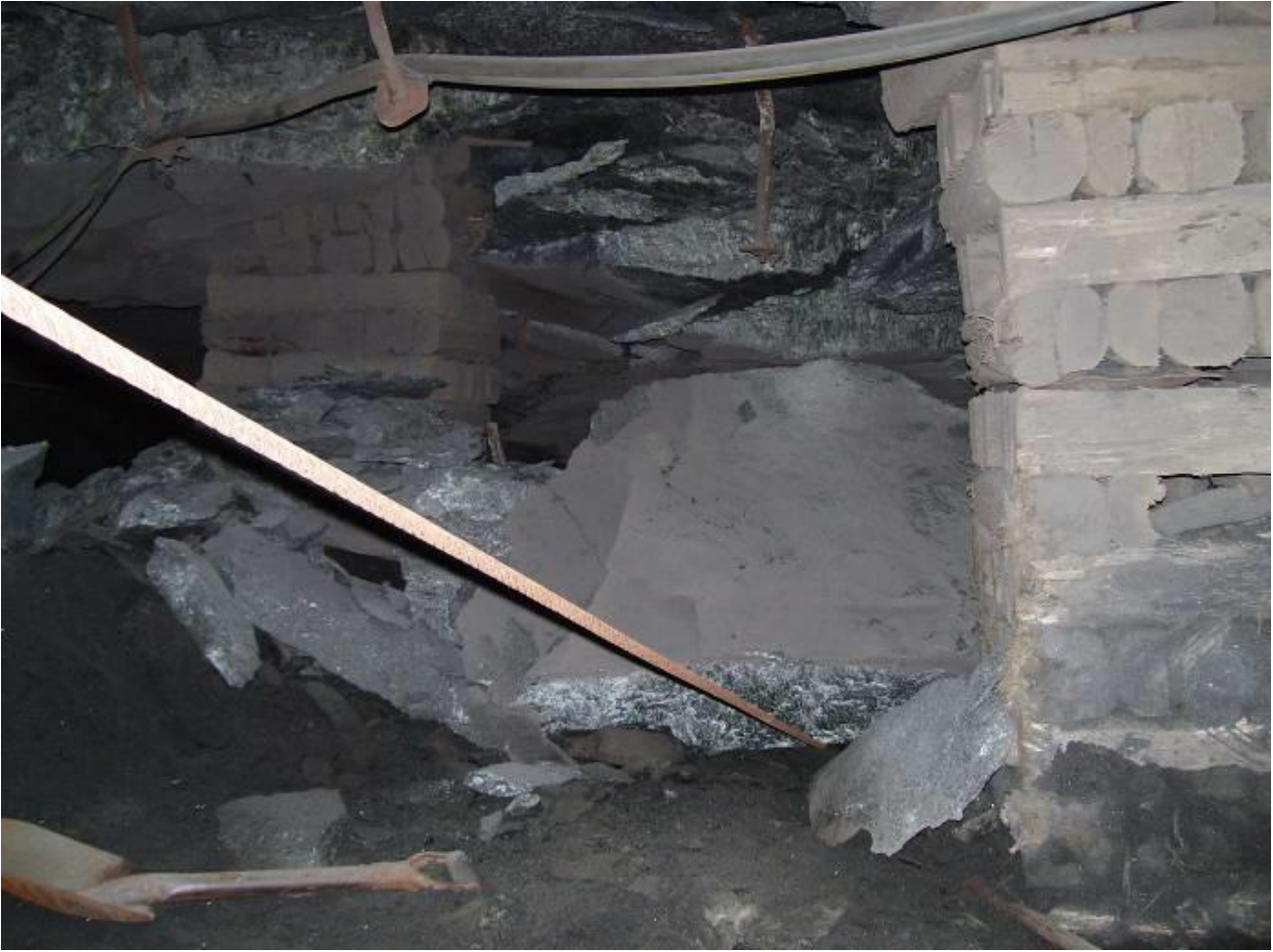

**Figure S8.** Collapse in Panel 7N on 04 April 2007.